\documentclass{aastex}

\slugcomment{Submitted to The Astrophysical Journal}

\begin{document}

\title{The frequency of nuclear star-formation in Seyfert\,2 galaxies}

\author{Thaisa Storchi-Bergmann\altaffilmark{1}$^,$\altaffilmark{2},
Daniel Raimann\altaffilmark{2}$^,$\altaffilmark{3}, Eduardo L. D. Bica\altaffilmark{2} and Henrique A. Fraquelli\altaffilmark{2}$^,$\altaffilmark{3}}
\affil{Instituto de F\'\i sica -- UFRGS,
	CP\,15051, CEP\,91501-970,
	Porto Alegre, RS, Brasil}
\email{thaisa@if.ufrgs.br}





\altaffiltext{1}{Visiting Astronomer at the Cerro Tololo 
Interamerican Observatory, operated by the Association of Universities for 
Research in Astronomy, Inc. under contract with the 
National Science Foundation.}

\altaffiltext{2}{e-mail addresses: thaisa@if.ufrgs.br, raimann@if.ufrgs.br, bica@if.ufrgs.br, ico@if.ufrgs.br}

\altaffiltext{3}{CNPq Fellow}

\begin{abstract}

We investigate the detectability of starburst signatures in the
nuclear spectrum of Seyfert\,2 galaxies by constructing 
spectral models in the wavelength range $\lambda\lambda$3500--4100\AA, 
combining the spectrum of a bulge population (of age $\approx$10~Gyr) 
with that of younger stellar populations, spanning ages from 
$\approx$3\,Myr to 1\,Gyr.
The major constraints in the analysis are:
(i) the continuum ratio 3660\AA/4020\AA, which efficiently discriminates 
between models combining a bulge spectrum with a stellar 
population younger than $\approx$50~Myr and those with older
stellar populations; (ii) the presence of the Balmer lines H8, H9 and 
H10 in absorption, which are unambiguous signatures of 
stellar populations with ages in the range 10\,Myr--1\,Gyr
for the relevant metallicities. Their detectability
depends both on the age of the young component and
on its contribution to the total flux relative to that of the bulge.

We also construct models combining the bulge template with a power-law 
(PL) continuum, which is observed in some Seyfert\,2's in polarized light,
contributing with typically 10--40\% of the flux at $\lambda$4020\AA.
We conclude that such continuum cannot be distinguished from that of
a very young stellar population (age$\le$10\,Myr), contributing 
with less than $\approx$0.02\% of the mass of the bulge.

The models are compared with nuclear spectra ---
corresponding to a radius of 200--300\,pc at the galaxy --- of 20  Seyfert\,2
galaxies, in which we specifically look for the signatures above
of young to intermediate age stellar populations. We find them
in ten galaxies, thus 50\% of the sample. But only in six cases (30\%
of the sample) they can be attributed to young stars (age $<$500\,Myr):
Mrk\,1210, ESO\,362-G8, NGC\,5135, NGC\,5643, NGC\,7130 and NGC\,7582.
In the remaining four cases, the signatures are due to intermediate
age stars ($\approx$1\,Gyr). 

We find a tendency for the young stars to be found more frequently
among the late type Seyfert's. This tendency is supported 
by a comparison between the equivalent widths (W) of absorption lines 
of the nuclear spectra of the  Seyfert\,2's with those of
normal galaxies of the same Hubble type: for the late-type 
(Sb or later), the W values of the Seyfert's are within the 
observed range of the normal galaxies, while the W values are
lower than those of the normal galaxies for 7 out of the 11 
early-type galaxies (S0 and Sa).


\end{abstract}

\section{Introduction}

The connection between star-formation and nuclear activity in galaxies
has been the subject of a number of recent studies. On the theoretical
side, models such as those of Perry \& Dyson (1985) and Norman \&
Scoville (1988) propose that a nuclear young stellar cluster is the reservoir
of fuel for the AGN in the nucleus.
 
On the observational side, Terlevich, Diaz \& Terlevich\,(1990) 
have argued that for a number of  Seyfert\,2 galaxies, 
circumnuclear starbursts are necessary
to explain the strength of the Ca\,II triplet (at $\lambda\approx$8500\AA) 
in absorption. Starbursts
have also been proposed as the source of the blue unpolarized continuum
observed in  Seyfert\,2 galaxies (FC2, Tran 1995a,b,c) by Cid Fernandes
\& Terlevich (1992, 1995). Heckman et al. (1995) have
proposed the same on the basis of the IUE spectra of the brightest 
 Seyfert\,2 galaxies in the UV. Detailed studies of individual cases
have been performed by Heckman et al. (1997) who have shown  HST UV
images and spectra of the circumnuclear
starburst around Mrk\,477, and Gonz\'alez Delgado et al. (1998),
who has analyzed similar data for the starbursts 
around the  Seyfert\,2 nuclei of 
NGC\,7130, NGC\,5135 and IC\,3639.
The latter studies have shown the most unambiguous signatures of massive
starbursts around these nuclei, but have been performed for only
4 galaxies.  

Our main goal in this work is to extend to a larger sample the search
of such nuclear starbursts  signatures in AGN's.
Recent works (e.g. Gonz\'alez Delgado, Leitherer \& Heckman 1999)
have pointed out that some of the best spectral features to date
starbursts are found in the blue-near UV spectral region (3400--4200\AA),
the strongest being the higher order Balmer absorption lines
(hereafter HOBL). These features seem indeed to be present in the spectra
of most starburst galaxies, as can be observed in the sample 
of Storchi-Bergmann, Kinney \& Challis (1995).
Bica, Alloin \& Schmitt (1994, hereafter BAS94) have
also pointed out the importance of this spectral range for the 
interpretation of composite stellar populations because of the 
signatures of the younger components (e.g. Balmer lines and Balmer
jump), and the 4000\AA\ break related to the old and intermediate age 
($\approx$1--5\,Gyr) populations. They studied near-UV/blue spectra 
of star clusters of all ages,
and analyzed equivalent widths of a number of features and continuum fluxes 
as a function of age and metallicity.
Fig. 1 of BAS94 is particularly illustrative of the
variation of the HOBL as a function of age and metallicity.

In order to look for starbursts in Seyfert's, one has to take into account
the contribution of the underlying bulge population. Simulations of
starburst spectra superimposed on older stellar populations, for the
spectral range 3700--9600\AA\ were carried out by Bica, Alloin \& Schmidt
(1990, hereafter BAS90; see also Schmidt, Alloin \& Bica 1995). 
For 3 relative amounts of mass stocked in star formation events,
they combined a bulge spectrum with spectra
of star cluster templates of different ages to simulate the evolution
of equivalent widths and continuum fluxes. 
In the present study,
we construct similar models in the range 3600--4100\AA\ which is
particularly important for the Seyfert~2 issue. The present models
represent an improvement relative to BAS90 in the sense that the templates
incorporate the high S/N near-UV range constructed from the star
cluster spectra of BAS94.

We then apply the results of the above simulations to the nuclear spectra
of 20  Seyfert\,2 galaxies in order to quantify the contribution of a possible nuclear starburst to the spectrum. The data used here are the nuclear
spectra of the 20  Seyfert\,2's of the sample of Cid Fernandes, Schmitt
\& Storchi-Bergmann (1998). Schmitt, Storchi-Bergmann \& 
Cid Fernandes (1999) have performed a spectral synthesis
using these spectra, but most features used in the synthesis have
$\lambda>$ 4000\AA. Our goal here is to take a close look at the
near-UV region. The aperture used in the extraction of the nuclear spectra
(2\arcsec$\times$2\arcsec) corresponds at the galaxy to regions
within a radius 200--300\,pc from the nucleus, which are
of the order of the nuclear starburst sizes investigated
in detail by Gonz\'alez Delgado et al. (1998).

The second goal of the present study
is to compare the results obtained for the Seyferts
with those for normal galaxies of the same Hubble type.
Cid Fernandes et al. (1998) have shown that in many  Seyfert\,2's
the blue nuclear continuum seems to be absent, when one takes into
account the surrounding stellar population. In other words, they
have found that the nuclear stellar population is varied, and
usually differs from that of an elliptical galaxy, as adopted in
early works. The blue continuum artificially appears when one
subtracts an elliptical galaxy template from the nuclear spectrum
of a galaxy containing some contribution of younger stars (Storchi-Bergmann,
Cid Fernandes \& Schmitt 1998). 
The spectral syntheses of Schmitt et al. (1999) have
further shown that the main difference between the nuclear
stellar population of Seyfert\,2's and that of an elliptical galaxy
is an excess contribution of a $\approx$100\,Myr stellar population.
These findings point to FC2 as
due to young stars in several cases, but another important question is whether
the contribution of young stars in Seyfert nuclei is 
larger than that in normal galaxies of the same
Hubble type. Since Seyfert nuclei as a rule occur
in spiral galaxies, some contribution of young stars is expected.
In order to access this aspect, it is necessary to compare
the nuclear stellar population characteristics of the Seyfert's with
those of normal galaxies. We thus assume the spectral 
study of Bica \& Alloin (1987)
as representative of the results obtained for normal galaxies to
make such a comparison.

The paper is organized as follows: in Section 2 we discuss the
near-UV features and adopted templates; in Sec. 3 we present the 
spectral models; in Sec. 4 we compare the nuclear  Seyfert\,2
spectra with the models; in Sec. 5 we compare our results
with those of similar studies;  in Sec. 6 we
discuss the degeneracy problem between a featureless power-law 
and a very young starburst continuum; in Sec. 7 we compare the
results for the Seyfert's with those for
normal galaxies of the same Hubble type, and finally in Sec. 8
we present our conclusions.

\section{Near-UV features and adopted templates}

The spectral features in the near-UV for a red (old/intermediate age) stellar population are markedly different from those of younger stellar populations. This can be observed in Fig. 1 (see also Fig. 2 of BAS94), where we show the stellar population templates used in this blue-violet study. 
The star cluster and
early-type galaxy bulge (E2E5 of BAS94) templates used here
incorporate observations from near-UV and visible (Bica \& Alloin
1986, 1987) domains. 
The templates also extend further to the UV and near-IR domains, 
which can be used as additional
constraints in the analysis. A detailed description of these templates
is given by Bica and collaborators in Section 4.1 of Leitherer et al.
(1996). To represent a 3\,Myr stellar population we adopt the integrated
spectrum of 30 Doradus (BAS94, Bica \& Alloin 1986).

In the old stellar population template, typical of elliptical galaxies or
of the bulge of spirals, the main absorption features are the
CaII K\,($\lambda$3933\AA) and H\,($\lambda$3970\AA) lines and a blend of 
absorptions due to CN, MgI, Si\,I and  Fe\,I. For the younger stellar populations, with ages between 10\,Myr and 1\,Gyr, the main features are,
apart from the CaII\,K and H lines with varying strengths, 
the high order Balmer lines
in absorption, in particular H$\delta$($\lambda$4102\AA), 
H$\epsilon$ ($\lambda$3970\AA),
H8($\lambda$3889\AA), H9($\lambda$3835\AA) and H10($\lambda$3797\AA).
Detection of the latter lines in a spectrum is thus a strong
signature of the presence of a blue (young) stellar population, except
when the burst is younger than 5\,Myrs, in which case these 
absorptions are filled by emission.

\section{Models}

Following BAS90, we have used the template bulge 
spectra in combination with
varying proportions of the younger templates in order to 
build simple model spectra representative of composite populations.
We have combined the bulge component with
0.1\%, 1\% and 10\% mass contributions locked in the younger components,
to construct the models shown in Figs. 2, 3 and 4. The adopted
mass-to-light ratios in V (BAS90) for the templates 
representing bursts of different ages, 
together with the relative luminosities at $\lambda$4020\AA\ 
L$_{\lambda4020}$ for 
bursts of equal masses, are listed in Table 1. This table can be used
to find the contribution in flux at $\lambda$4020\AA\ corresponding to a given
mass proportion. For example, in
order to construct the model combining the bulge with
0.1\% in mass of the 10\,Myr population, we have multiplied
L$_{\lambda4020}$ of the 10\,Myr template by 0.1\% to obtain 3.354,
which is the normalization factor of the template at $\lambda$4020\AA. 
Subsequently it is added to the bulge template normalized to 1 at this
same wavelength. Notice that for equal masses, the 10\,Myr cluster L$_{\lambda4020}$ value is higher than that of the 
3\,Myr cluster embedded in the
HII region because of the appearance of supergiants.

By examining the characteristic features of the young
stars in Figs. 2, 3, and 4, in particular the absorption 
lines H8, H9 and H10 (the HOBL) -- which
are the ones less affected by emission in Seyfert's, see below --
it can be concluded that: (i) if the starburst is very young,
with ages of a few Myrs, the HOBL are filled by emission;
in this case, the HII region emission spectrum may be used as constraint;
(ii) for a small starburst, so
that its mass is $\approx$0.1\% of that of the bulge, the HOBL
can be detected for young starbursts with
ages from 10 to 50\,Myrs; (iii) for a starburst corresponding
to 1\% of the mass of the old stars, the above features can be
observed for ages from 10 to 500\,Myrs; (iv) for a stronger
starburst, corresponding to 10\% of the mass of the bulge, the 
features are observed for ages from 10\,Myrs up to 1 Gyr. 

In order to quantitatively characterize stellar populations, 
BAS94 have proposed the use of a few continuum points, 
whereby a continuum is traced (connecting these points
using straight lines) to measure the 
equivalent widths (hereafter represented by W)
of  a number of features. However, in Seyferts,
many of these features are filled by emission lines and cannot be used. 
By comparing the above stellar population templates with typical Seyfert
spectra, we conclude that the windows which are usually free
from emission-lines are $\lambda\lambda$3810--3822\AA (centered on
a continuum point), $\lambda\lambda$3822--3858\AA (centered on H9)
and $\lambda\lambda$3908--3952\AA (centered on the CaII\,K line).
We have thus followed the BAS90's method to construct the continuum,
and measured the W's within the three windows above, hereafter
identified as W$_C$, W$_{H9}$ and W$_{CaK}$, respectively.
Fig. 5 illustrates the continuum and windows used for three  Seyfert\,2's with
distinct near-UV spectra. 

The near-UV W's and continuum ratio $\lambda$3660/$\lambda$4020
(hereafter CR) were  measured for the models. We show in 
Fig. 6 CR, W$_C$ and W$_{Ca\,II}$ 
as a function of the young component's age, 
for the three proportions in mass above. 
It can be concluded that CR (F$_{3660}/F_{4020}$) is a powerful 
star-formation tracer for very young stellar populations, 
varying from 1.4 for a 3\,Myr stellar population 
down to 0.6 for 50 Myrs or older. 
W$_C$ and W$_{CaIIK}$ are better age 
indicators for older bursts, but can trace
stellar populations of all ages, from 3\,Myr to 1\,Gyr. 
We recall that blue-violet metal lines in composite spectra are age indicators
because of the dilution effects caused by hot main sequence stars 
(Bica \& Alloin 1986). Regarding H9, we conclude that its profile,
together with those of the other HOBL are better age indicators than W$_{H9}$. 

A number of  Seyfert\,2 galaxies present a polarized blue continuum 
with a power-law spectrum (Tran 1995a,b,c; Storchi-Bergmann et al. 1998).
We have thus also constructed a second set of 
models combining the bulge template and
a power-law $F_{\nu}\propto \nu^{-1.5}$(hereafter PL), which is typical 
of the polarized continuum found by Tran (1995a,b,c).
As the mass proportions used above have no meaning for the PL,
we have used combinations with varying
proportions of the PL in flux at $\lambda$4020\AA.

The CR, W$_C$ and W$_{Ca\,II K}$ measured 
for the models combining the bulge template with the PL
are plotted in Fig. 7, as a function of the PL
percent contribution in flux at $\lambda$4020\AA. As a comparison,
we also plot the W values from models combining a bulge with
the same flux contribution from a 10\,Myr stellar population and
a 100\,Myr population. 
Combination with older populations produce similar CR values as
the latter, as can be seen in Fig. 6.  
Fig. 7 shows that CR is the most powerful
discriminator between a PL and bursts of $\approx$100\,Myr or older.
On the other hand, it shows also that a PL is hardly distinguishable 
from a 10\,Myr burst on the basis of the CR and W
values, a problem which has been already pointed out by several 
authors and by ourselves in Schmitt et al. (1999). 
We discuss this problem further in Sec. 6.   

\section{Comparison with Seyferts}

The Seyfert sample used in this work is the same as that of
Schmitt et al. (1999, hereafter SSC99) except for one radio-galaxy
(with had a low signal-to-noise ratio spectrum in the near-UV),
and consists of the 20  Seyfert\,2's
and 3 radio-galaxies from the larger sample of Cid Fernandes et al. (1998).
As in SSC99, we use  here only the nuclear spectra, 
extracted using windows of 2$\times$2\arcsec,
which correspond at the galaxy to regions of typical diameters
of a few hundred parsecs. 

These spectra are plotted in Figs. 8
to 12, separated according to the Hubble types.
Here we have made use of the updated morphological classifications 
of our sample of  Seyfert\,2's in the Malkan et al. (1998) HST optical
imaging survey, due to the better spatial resolution
and dynamical range of their images as compared with
those upon which the previous classifications (RC3, de Vaucouleurs et
al. 1991) are based. We found an interesting result: out of the 20 galaxies
in our sample, 10 have new classifications when compared with those
of RC3, {\it always to later Hubble types}, as shown in Table 3.
In particular, two of the galaxies with the most obvious 
signatures of recent star formation,
NGC\,5135 and NGC\,7130 (Gonz\'alez Delgado et al. 1998), previously
classified as Sab and Sa, respectively, have been re-classified as
Sc and Sd by Malkan et al. (1998).

The galaxies CR, W's and scales (pc/arcsec) are listed in Table 2.
Typical errors are $\sim$0.01 in CR and 
$\le$1\AA\ in the W's.
In Cid Fernandes et al. (1998) we have noticed
that the nuclear spectra of these galaxies were in most cases redder than the
extranuclear neighboring spectra, a result that we 
have attributed to reddening.
Indeed, in the spectral syntheses performed by SSC99,
they usually found significant reddenings 
(0.1$\le$E(B--V)$\le$0.60). We have thus corrected the nuclear spectra by the 
reddening (Seaton 1979) found by SSC99 before tracing the continuum and
measuring the near-UV W's and CR.
Notice however that the reddening
corrections are not critical for CR due to the proximity of 
the continuum wavelengths $\lambda$3660\AA\
and $\lambda$4020\AA. 

We now compare the near-UV nuclear spectra of our sample
with the synthetic spectra derived from
the CR and W's listed in Table 2, 
using the model values from Figs. 6 and 7.
  
As the templates have lower spectral resolution ($\approx$15\AA)
than the Seyfert spectra ($\approx$5\AA), we present here
the models compared with the Seyfert spectra smoothed to better match
the template resolution. 
The smoothing may cause loss of information,
particularly when there are faint emission lines superimposed on the HOBL. 
We have thus always checked for these emission lines at full resolution 
and have not included in the fit the absorption features affected by 
line emission. The most frequent case is the contamination of 
the CaII\,H + H$\epsilon$ 
absorption by NeIII$\lambda$3968 + H$\epsilon$ emission.

Our goal with these comparisons is not a perfect match of the
nuclear spectrum, but to identify the unambiguous
signatures of the different components to the nuclear spectrum,
looking in particular for the characteristic
star-formation features HOBL (H8, H9 and H10 in absorption),
using the simple models as a guide. With this approach, we also check
if these simple models 
are a good representation of the nuclear stellar population.
In applying the models of Figs. 6 and 7, 
it can be noted that the CR's and
W's are compatible in several cases with more than one model. 
In these cases, the
adopted model is the one which gives the best fit to the overall spectrum.
The quality of the fit is inspected in regions free from emission lines,
and the best fit is the one which gives the smaller residuals 
between the observed and model spectra in these regions. 

 As pointed out above, it is not possible to distinguish
a PL continuum from that of a template of age 10\,Myr or younger,
for flux contributions smaller than 40\% at $\lambda$4020\AA. 
When such a continuum is needed we will call it PL/YS, meaning
power-law or young stars. The nature of this component is further
discussed in Sec. 6.


If the  Seyfert\,2 has strong Balmer emission-lines, the Balmer continuum
in emission may be important in the near-UV region, diluting the W values
calculated above. From the emission-line fluxes, we found that the Balmer
continuum only contributes significantly (with more than 5\% to the flux) 
for $\lambda<3646$\AA, for the galaxies IC\,1816, MCG\,05-27-013, Mrk\,348, Mrk\,573 and Mrk\,1210. In these cases we have considered also 
the contribution of the Balmer continuum, calculated 
as in Osterbrock (1989),  and normalized according to the
fluxes of the Balmer emission lines.



\subsection{Elliptical galaxies}

In order to verify the applicability of the templates as a basis
to synthesize the Seyfert spectra, which were observed with the
CTIO 4m Blanco telescope (Cid Fernandes et al. 1998), 
we have first applied the models 
to the normal elliptical galaxy IC\,4889, observed with the same
telescope, as a test to the method. 
In addition, we apply the models to three radio elliptical
galaxies. The spectra and models are illustrated in Fig. 13.

\subsubsection{The normal elliptical IC\,4889}

The near-UV W's and CR agree very well with those
of the bulge template, and indicate no need
of bluer components, as expected. Fig. 13a illustrates the
observed spectrum as compared with the bulge template, showing
a very good match of the observed spectrum to the template,
indicating that we can use the templates constructed from the cluster 
spectra to synthesize the Seyfert nuclear spectra. Small differences
are apparent in the CaII\,K and H lines, which are deeper in the galaxy
spectrum, even though the W's are the same as in the bulge template. 
We attribute this effect to a residual difference in spectral resolution 
between our spectrum and the bulge template.

\subsubsection{3C\,33}

This is a radio galaxy with near-UV W values smaller than those 
of the bulge template, indicating the presence of some contribution of
a blue component. The best fit for the spectrum is obtained with the combination
of a bulge plus 10\% in mass of a population
of 1\,Gyr, although the nuclear spectrum of the galaxy is still
somewhat bluer for $\lambda<$3660\AA. 
An improved fit to the blue end of the spectrum
is obtained by adding a 5\% contribution in flux at $\lambda$4020
of the PL/YS component. The HOBL are at the detection limit (Fig. 13b). 
NeIII$\lambda$3968 + H$\epsilon$ emission can be observed filling
the CaII\,H + H$\epsilon$ absorption.

\subsubsection{PKS\,0349-27}

This radio galaxy has a small dilution of the W's when compared
with the elliptical template values. The overall spectral
distribution is best reproduced by the combination of the bulge with 10\%
flux contribution of the PL/YS component at $\lambda$4020\AA\,(Fig. 13c). 
The difference in the depth of the Ca\,II\,K line is similar to
that observed in IC\,4889 and thus due to
the better spectral resolution of the radio galaxy spectrum relative to the template. It is not possible to identify the HOBL.
NeIII$\lambda$3968 + H$\epsilon$ emission is filling the 
CaII\,H + H$\epsilon$ absorption.

\subsubsection{PKS\,0634-20}

Here the case is similar to the one above, with a somewhat larger
dilution in the continuum window and somewhat larger CR.
The CR value indicates a 15\% PL/YS contribution, in agreement
with the W values and overall spectral distribution
(Fig. 13d). The same remark above about the CaII\,K line applies here. 
It is not possible to identify the HOBL.
There is contamination of the CaII\,H + H$\epsilon$ 
absorption by NeIII$\lambda$3968 + H$\epsilon$ emission.

In summary, the near-UV spectrum of the three radio galaxies, 
when compared with normal ellipticals,
shows a systematic need of a small (10-20\% in flux at
$\lambda$4020\AA) contribution of a blue component.
For the 2 PKS sources, this component is well reproduced by a
PL/YS continuum. For 3C\,33 it is necessary to add
also the contribution of a 1\,Gyr stellar population
with $\approx$10\% of the bulge component mass. 
A possible interpretation for such large intermediate age
contribution would be the cannibalism of a small spiral galaxy or
magellan irregular, as is the case, for example, of the nearby radio
galaxy Centaurus A (e.g. Storchi-Bergmann et al. 1997).
 
\subsection{S0 and S0a}

\subsubsection{NGC\,1358}

In this galaxy the  near-UV nuclear features and spectral distribution are well
represented by the bulge template alone (Fig. 14a). There is some 
contamination of the CaII\,H + H$\epsilon$ 
absorption by NeIII$\lambda$3968 + H$\epsilon$ emission.

\subsubsection{ NGC\,3081}

The CR indicates the presence of a PL/YS component contributing 
with $\approx$25\% of the flux at $\lambda$4020\AA.
The near-UV W's and overall spectrum are also well reproduced by
the above combination (Fig. 14b). HOBL are not detected.
NeIII$\lambda$3968 + H$\epsilon$ emission can be observed filling
the CaII\,H + H$\epsilon$ absorption.


\subsubsection{Mrk\,348} 

CR is 0.72 for this galaxy, thus indicating the presence of
a PL/YS component contributing with 30\% in flux at $\lambda$4020\AA, and 
in approximate agreement with the W values. Fig. 14c shows also that
the overall spectrum is well reproduced by this combination, taking
into account that there is some emission in H9 and probably also
in H10, making it impossible to detect the HOBL in absorption, if any. 
NeIII$\lambda$3968 + H$\epsilon$ emission completely dominates over
the CaII\,H + H$\epsilon$ absorption.


\subsubsection{Mrk\,573}

H9 is filled by emission. The CR indicates the presence of a PL/YS component
contributing with $\approx$20\% in flux at $\lambda$4020\AA.
The other two W's and overall continuum are also well 
reproduced by this combination (Fig. 14d). The small discrepancy
in the CaII K line profile is explained as for the radio-galaxies
above.  HOBL are not detected. 
NeIII$\lambda$3968 + H$\epsilon$ emission completely dominates over
the CaII\,H + H$\epsilon$ absorption.


\subsubsection{IRAS\,11215-2806}

 This galaxy shows significant dilution in the near-UV W's, with
the CaII\,K line profile better reproduced
by a combination of a bulge plus 20\% mass contribution  of a 1\,Gyr component
-- an additional model we had to construct increasing the contribution 
of the intermediate age population to fit the spectrum of
this galaxy (Fig. 14e). The HOBL can be identified.
There is some contamination of the CaII\,H + H$\epsilon$ absorption
by NeIII$\lambda$3968 + H$\epsilon$ emission.


\subsubsection{Fairall\,316}

The W values are very similar to those of the old bulge template, and Fig. 14f
shows that the latter is indeed a good representation of the nuclear
spectrum of this galaxy.
There is some contamination of the CaII\,H + H$\epsilon$ absorption
by NeIII$\lambda$3968 + H$\epsilon$ emission.

\subsubsection{ESO\,417-G6}

The near-UV W's, CR and overall spectrum of this galaxy are best
reproduced by the combination of a bulge template plus 10\% mass contribution
of a 1\,Gyr stellar population (Fig. 14g). 
The HOBL are at the detection limit, but can be identified because
there seems to be no contamination by H9 and H10 emission.
There is some contamination of the CaII\,H + H$\epsilon$ absorption
by NeIII$\lambda$3968 + H$\epsilon$ emission.

In summary, of the seven nuclear spectra of S0 Seyfert's 2, two can be 
reproduced by a bulge stellar population and three are better reproduced by a combination of the bulge template with the PL/YS component.
The HOBL signatures can be observed in IRAS\,11215-2806
and ESO\,417-G6 due to a large contribution of an intermediate age
1\,Gyr stellar population.

\subsection{Sa}

\subsubsection{ Mrk\,1210}


 CR is 0.83 for this galaxy suggesting a PL/YS component contributing with 
$\approx$50\% of the flux at $\lambda$4020, which is in agreement
with the value obtained from W$_{CaIIK}$. The other two W values
are contaminated by emission lines, which precludes the detection
of the HOBL. Storchi-Bergmann et al. (1998) have shown that the
extranuclear spectrum is dominated by an intermediate age population,
which can be represented by our combination of a bulge plus 1\% mass
contribution of a 500\,Myr stellar population. They have also shown
that the nuclear spectrum could be well reproduced by the latter 
population plus a very young starburst, as evidenced by the  
Wolf-Rayet features (Storchi-Bergmann et al. 1998,
Cid Fernandes et al. 1999). We have then constructed a 
model combining 70\% in flux at $\lambda$4020
of the extranuclear population plus 30\% of the 3\,Myr population
to represent the nuclear spectrum of this galaxy, which is
shown together with the observed nuclear spectrum in Fig. 15a. 

The emission-lines in the nuclear spectrum are
stronger than in the model, indicating that the observations are consistent
with the contribution of a 3\,Myr stellar population but 
suggesting also that another source of
continuum -- the AGN continuum, is necessary to ionize the gas.

\subsubsection{CGCG\,420-015}

The near-UV spectrum is
best reproduced by a mixture of bulge plus 10\% in mass of
an intermediate age (1\,Gyr) stellar population (Fig. 15b).
The spectrum is somewhat noisy in the region, 
which together with some contamination by emission lines makes difficult
the identification of the HOBL.

\subsubsection{ IC\,1816}


The CR of 0.66 suggests $\approx$15\% 
contribution of a PL/YS continuum. The W's suggest additional
dilution which can be provided by an intermediate age population.
We thus show in Fig. 15c two models: the bulge plus
15\% flux contribution at $\lambda$4020 of the PL/YS component, 
and the improved fit provided by
combining the bulge with 10\% in mass of a 1\,Gyr stellar population
before the combination with the PL/YS. The strong emission precludes
a firm identification of HOBL.


\subsubsection{ESO362-G8}

The nuclear spectrum of this galaxy unambiguously shows all HOBL
in absorption. The near-UV W's are well reproduced 
by the bulge template combined with 10\% mass  contribution of a 
stellar population with age between 100 and 500\,Myrs,
with the metal lines of the  Seyfert\,2 galaxy somewhat deeper
most probably because of the lower metallicity of the star clusters used
to construct the young stellar templates. The CR measured for this galaxy was
0.35, much lower than the lowest template value (0.55), suggesting that the continuum was still reddened. We have thus corrected 
the continuum of this galaxy by an additional E(B-V)=1, which
brought the CR value to 0.52 (about the minimum obtained for the
sample), in agreement with the model selected from the W's (Fig. 15d).
This very high reddening is consistent with the dust lane observed
crossing the nuclear region (e.g. Malkan et al. 1998), and is in
agreement with the E(B-V) values obtained from emission lines
(Fraquelli, Storchi-Bergmann \& Binette 2000).

In summary, among the 4 Sa's, there is one 
unambiguous case of a relatively evolved  nuclear starburst, 
ESO\,362-G8, clearly showing the HOBL in absorption. 
In addition, Mrk\,1210 has a very young burst, 
for which the HOBL are filled with emission. For CGCG\,420-015,
and IC\,1816 there is some contribution of intermediate age population and
for the latter, there is also evidence of a PL/YS continuum.

\subsection{Sab, Sb and Sbc}

\subsubsection{NGC\,1386}

The W's, CR and spectrum are best reproduced by the bulge 
template plus 10\% im mass of
a 1\,Gyr population. This fit is shown in Fig. 16a, where the
the HOBL can be observed. The NeIII$\lambda$3968 + H$\epsilon$ emission
are almost filling the CaII\,H + H$\epsilon$ absorption.


\subsubsection{NGC\,6890}

The CR of 0.7 indicates the presence of a PL/YS continuum
contributing with $\approx$30\% to the flux at $\lambda$4020,
in agreement also with the W values. Some contribution
of an intermediate population is also possible, and a
model including 10\% mass contribution of a 1\,Gyr 
stellar population, combined with 20\% in flux
of a PL gives a slightly improved fit in
the H9--H10 region, as shown in Fig. 16b.
NeIII$\lambda$3968 + H$\epsilon$ emission
are filling the CaII\,H + H$\epsilon$ absorption.


\subsubsection{NGC\,7582}

This galaxy is well known from the emission-lines (diagnostic diagrams)
to present a composite  Seyfert\,2 + Starburst spectrum.
The HOBL are easily seen (Fig. 16c). The near-UV W's, CR and
spectral distribution are well represented by a bulge template plus
1\% mass contribution of a 50\,Myr stellar population.
Again, we attribute the poor fit of the CaII\,K line 
mostly to the lower metallicity of the young templates 
as compared to the nuclear region of the galaxy.
There is contamination of the CaII\,H + H$\epsilon$ absorption
by NeIII$\lambda$3968 + H$\epsilon$ emission.

\subsubsection{Mrk\,607}

The near-UV W's, CR and spectrum are best reproduced by a bulge plus
10\% mass contribution of a 1\,Gyr stellar population (Fig. 16d).
The spectrum is somewhat noisy in the region, and thus the HOBL cannot 
be unambiguously identified. NeIII$\lambda$3968 + H$\epsilon$ emission
are filling the CaII\,H + H$\epsilon$ absorption.



\subsubsection{MCG-05-27-013}

CR is 0.7 for this galaxy suggesting a 20\% contribution of a PL/YS component.
The spectrum is dominated by line emission, including in the
high order Balmer lines and the blue end of the spectrum is noisy. It is not possible to identify the HOBL in absorption (Fig. 16e).



In summary, for the five Sab, Sb and Sbc's, NGC\,7582
presents the  HOBL signatures of a nuclear starburst, NGC\,1386,
Mrk\,607 and possibly also NGC\,6890 
have 10\% contribution in mass of intermediate age 1\,Gyr stars and 
the CR of MCG-05-27-013 and NGC\,6890 suggest the presence  of a PL/YS
component contributing with 20\% in flux at $\lambda$4020\AA.

\subsection{Sc and Sd}

\subsubsection{NGC\,5135}


The HOBL are clear in the spectrum.
The W's, CR and other features of the 
spectrum can be approximately reproduced 
by the combination of 50\% each in flux at $\lambda$4020
of two of our models: the first is the bulge
template combined with 1\% in mass of a  10\,Myrs stellar population,
and the second is the bulge template combined with 1\%
in mass of a 100\,Myrs stellar population (Fig. 17a).
There is some line emission of H$\epsilon$+NeIII$\lambda$3968, 
H8 and possibly H9.

The starburst in this galaxy has been extensively studied by Gonz\'alez Delgado
et al. (1998) and Gonz\'alez Delgado, Heckman \& Leitherer (2000), 
hereafter GD98 and GD00.
GD98  have shown that the UV spectrum presents clear signatures of
O and B stars, estimating an age between 3 and 5 Myr for the burst,
while GD00 concluded that there is also a similar contribution in 
flux from an intermediate age population and a small contribution of
an old component. Our model is consistent with the latter mixture.

\subsubsection{ NGC\,5643}

The near-UV W's, CR and spectrum can be 
best reproduced by the combination
of the bulge template plus 1\% in mass of a 100\,Myr stellar 
population, except for the Ca\,II K line, mostly due to 
the lower metallicity of the young template, as discussed previously. 
The HOBL can be identified (Fig. 17b).
NeIII$\lambda$3968 + H$\epsilon$ emission
are almost filling the CaII\,H + H$\epsilon$ absorption.

\subsubsection{ NGC\,6300}

 The near-UV W's, CR and spectral distribution are best reproduced
 by a mixture of a bulge
template plus  10\% mass contribution of a 1\,Gyr stellar population.
The model is compared with the data in Fig. 17c, where the
HOBL are at the limit of detection.


\subsubsection{ NGC\,7130}

Fig 17d shows the clear HOBL in this spectrum, another case
of nuclear starburst well studied by GD98 and GD00. The latter
authors have suggested the same population as that derived for NGC\,5135.
Although the W's support the same population, the bluer continuum 
of NGC\,7130 suggests a somewhat younger population, or a mixture
including a larger proportion of the younger components.
The model we show in the figure is the same composite model as
the one adopted for NGC\,5135, with a somewhat larger contribution
in flux at $\lambda$4020 of the bulge combined with the 10\,Myr 
stellar population: 75\%, while the bulge combined with 
the 100\,Myr population contributes with 25\% of the flux.

In summary, for the 4 Sc and Sd Seyfert's, three present
recent episodes of star formation in the nuclear region,
with the HOBL clearly visible in the spectra, while one presents
$\approx$10\% mass contribution of an intermediate age population.

\section{Comparison with other works}
 
The above results can be compared with those from the synthesis of
SSC99. There is good agreement for most galaxies
for which the HOBL are observed in the spectra: the young components
in our simple  models coincide, approximately, with 
the dominant young components obtained by SSC99.
One systematic difference seems to be the contribution of the
intermediate age components, found to be present in most cases by SSC99,
but only in approximately half of the sample here. We attribute
this difference to two factors: (1) the models of SSC99 were
not as strongly constrained in the blue end of the spectrum as in
the present work, provinding an optimized representaion of the data over
the espectral range $\lambda\lambda$3700--7000\AA;
(2) the lower metallicity of some of the
star clusters used to construct the 1\,Gyr template,
which is a mixture of spectra of LMC and Milky Way disk clusters,
while the synthesis performed by SSC99 is based on a grid of parameters
including the high metallicity end.

Gonz\'alez Delgado et al. (2000, GD00) have recently finished a
similar spectral study in which they investigate the age of the
stellar population at and around the nucleus of 
a sample of (also) 20  Seyfert\,2 galaxies.
They cover the spectral region $\lambda\lambda$3700-4400\AA\
at a similar spectral resolution to our observed spectra, 
but using models with the same resolution.
This better resolution in the modeling allows the
detection of the He\,I absorption (e.g. $\lambda$3819, 4387 adn 4922)
in some cases, providing a better dating of the starbursts.  
The higher spectral resolution also allows a more precise
evaluation of the emission-line contamination in the high
order Balmer lines.

GD00 find signatures of recent star formation
in the nuclear spectrum of 6 galaxies plus 3 or 4 cases in which
these signatures are found once the nebular Balmer emission lines
are subtracted. In two other cases they find significant contribution
of intermediate age stars. There are 5 galaxies in the sample
of GD00 in common with the present study: Mrk\,348, Mrk\,573, NGC\,1386,
NGC\,5135 and NGC\,7130. Similarly to what we have found, they conclude
that  Mrk\,348, Mrk\,573 and NGC\,1386 have a dominant old stellar population,
and that the near-UV nuclear spectrum of 
NGC\,5135 and NGC\,7130 is dominated by light from  young and intermediate
age stars. 

\section{The nature of the PL/YS continuum}

As pointed out above, there is a 
degeneracy between the power-law and starburst continua
for ages $\le$10\,Myrs. This occurs for mass contributions of the
young component much smaller than those in the models of Fig. 6: for example,
for a typical 30\% flux contribution of PL/YS at $\lambda$4020\AA,
if it is due to the continuum of a 10\,Myr stellar population, 
the corresponding mass contribution is only $\approx$0.015\% 
that of the bulge. This degeneracy is illustrated in Fig. 18, 
where we plot the combined spectrum of a bulge and a PL
contributing with 20\% and 40\% in flux at $\lambda$4020\AA, together
with the combined spectra of the bulge and a 10\,Myr
star cluster template, for the same proportions in flux as the PL,
and with the combined spectrum of the bulge and a 3\,Myr
star cluster contributing in flux with half the proportions above,
namely 10\% and 20\%. A young starburst in nature will most probably
present a spread in age, and the spectral distribution
will probably be more similar to a combination of the 3 and 10\,Myr
templates.

It can be observed that, for $\lambda\ge$3500\AA,
the combination with a PL can hardly be distinguished
from that with the young burst templates if their contribution 
to the flux is 20\%. The only constraint here could be the
strength of the emission lines.

Figure 18 shows that contributions in flux $\ge$40\% at $\lambda$4020\AA\
and good signal-to-noise (S/N) ratio spectra are necessary
to allow the distinction between a featureless PL and
a young starburst in terms of continuum features, such as the
HOBL. Additional constraints could be the emission-line strengths
and the UV spectrum slope. The UV spectrum can be observed rising much
more steeply for the 10\,Myr stellar population than for the PL.
Unfortunately our data do not extend enough to the
UV to allow the use of this constraint.

In our sample, the PL/YS contribution is always
smaller than or equal to 30\%, and it is thus not 
possible to determine its origin
from the near-UV features alone. In Storchi-Bergmann et al. (1998),
we have discussed the origin of the near-UV continuum of Mrk\,348 and Mrk\,573:
in the former, from the work of Tran (1995a,b,c), 
$\approx$10\% of the flux is due to scattered light, and thus
the remainder could be due to young stars. Population synthesis
has indeed shown that a mixture of a 10\,Myr and a 3\,Myr stellar
population can reproduce the continuum.
In Mrk\,573, at least
part of the blue continuum is also due to scattered light, as
revealed by the images of Pogge \& De Robertis (1993), but we have no
constraint on its value.

The only constraint we could try to use here
is the strength of the emission-lines; for example,
if the near-UV continuum were entirely due to a very young starburst
(the 3\,Myr one), the equivalent widths W$_{em}$'s 
of the Balmer emission lines in the nuclear spectra should be
similar to that of the model combining the bulge and the 3\,Myr
template. Most frequent in nature is however the case
in which there is some spread in age for the starburst so that
the blue continuum is also due to non-ionizing (ageing)  blue stars,
and the W$_{em}$'s should then be smaller than in the simple model
with the 3\,Myr template. 

We have measured the equivalent width
W$_{H\beta}$ of the nuclear H$\beta$ emission line
and found that it was smaller than in the simple model above for 
the 2 radio galaxies PKS\,0349-27, PKS\,0634-20 and for NGC\,6890,
and about the same order for NGC\,3081. In all other cases --
3C\,33, Mrk\,348, Mrk\,573, and MCG\,05-27-13,
the observed nuclear W$_{H\beta}$'s are larger than in the model. 
We can thus conclude that, if a starburst is
the origin of the blue continuum in the above galaxies, it is
an ageing one (age$\ge$10\,Myr) for the first three above, 
and could be a very young ($\approx$3\,Myr) in the other cases.
But we also remark that, although the blue continuum could be
due to young stars, the emission-line ratios in all cases (and even the
W's in the last three cases) require additional ionizing sources --
in other words, the starburst continuum cannot account alone
for the emission-line ratios (and emission-line 
luminosities in the last three cases).


\section{Comparison with ``normal'' galaxies of the same Hubble type}

A comparison can be made between the near-UV properties of the
Seyfert's above and non-Seyfert's of the same Hubble type.
Using the CaII\,K W's values from Bica \& Alloin (1987)
as representative of a sample of normal galaxies, we obtain the following
typical values, calculated as averages from approximately 30
galaxies of each Hubble type:
(i) for Sa's, W$_{CaII K}$=16.5$\pm$1.6\AA; (ii) for Sb's,
W$_{CaII K}$=13.8$\pm$4.2\AA\ and (iii) for Sc's, W$_{CaII K}$=9.9$\pm$4.9\AA.
Note that for later types, the spread of W's increases, simply reflecting
the variety of mixtures of old (bulge) stellar population and star-forming 
events in the central region of these galaxies.
For ellipticals and S0's, similar values to those of Sa galaxies are
obtained. 

From the 20 Seyfert's, 11 are early-type galaxies, classified as S0 or Sa;
in NGC\,1358 and Fairall\,316, W$_{CaII K}$ is
typical of early-type galaxies; in the other 9, the W's are smaller,
indicating the need of a blue continuum. This blue continuum
is clearly due to recent (age $<500$\,Myr) enhanced star-formation in ESO\,362-G8
and in Mrk\,1210, and to intermediate age stars in IRAS\,11215-2806,
and  ESO\,417-G6. For
Mrk\,348, Mrk\,573 and NGC\,3081, a PL/YS component is favored. 
This component is at least partially due to scattered light 
observed as polarized 
light in Mrk\,348 and Mrk\,573, but the unpolarized flux
could be partially 
originated in the continuum of a nuclear starburst younger than 10\,Myr.

Nine sample galaxies have later Hubble types, from Sb to Sd. 
Contrarily to the earlier types, the W$_{CaII K}$ values are within the range
observed for non-Seyfert's. Four of these galaxies
present recent (age $<$500\,Myr) episodes of star-formation in the nuclei
as revealed by the HOBL in NGC\,7582, NGC\,5643, 
NGC\,5135 and  NGC\,7130. NGC\,1386, NGC\,6300 and Mrk\,607 
are  cases in which there is enhanced contribution 
of an intermediate age population of $\approx$1\,Gyr.
MCG\,5-27-13 and NGC\,6890 are cases in which a PL/YS is necessary.

In summary, recent star formation episodes have been found in the
nuclei of 2 of the 11 early type Seyfert's and 4 of the 9 late type
Seyfert's, showing a tendency of these episodes to be found more frequently
in late-type Seyfert's.  




\section{Summary and Conclusions}

We have constructed models combining spectral distributions of
a typical bulge plus a young stellar population template
in the spectral range $\lambda\lambda$3600-4100\AA\ in order 
to investigate the detectability of
recent star-formation episodes in the nuclear region of Seyfert 
galaxies, which usually have prominent bulges.

The high order Balmer lines (HOBL) are
good indicators of the presence of young to intermediate
age populations as far as the mass contribution of these young components,
as compared with the bulge mass, is larger than 1\%
for ages of $\approx$100\,Myr, or larger than $\approx$0.1\%
for ages of  $\approx$10\,Myr. We
conclude in addition that the continuum ratio
CR=$\lambda$3660/$\lambda$4020 is an important discriminator
of very young stellar populations, as its value
can only be larger than CR=0.6 for models
including stellar populations younger than 50\,Myr.
If the flux contribution at $\lambda$4020 of these
very young populations is smaller than 40\% (corresponding to
0.02\% of the mass of the bulge for the 10\,Myr template), 
it is not possible to distinguish its spectral signatures (e.g., the HOBL) 
from a featureless power-law. Thus, the problem of degeneracy between the 
featureless AGN continuum and the continuum of a stellar population
of 10\,Myr or younger still remains at the above flux levels.

By comparing the nuclear spectrum of a sample of 20  Seyfert\,2's and
3 radio-galaxies with the models above, 
signatures of recent to intermediate age star-formation in the form of 
high-order Balmer absorption lines (HOBL) have been found
in 9  Seyfert\,2's. From previous analyses of the emission-line features,
enhanced recent star-formation has been found also in
Mrk\,1210. In summary, half of our Seyfert\,2 
sample show  signatures of young to intermediate stars.
In six cases (30\% of the sample), the starburst is
younger than 500\,Myrs: ESO\,362-G8, 
NGC\,7582, NGC\,5135, NGC\,5643 and
NGC\,7130. In the cases of ESO\,417-G6, IRAS\,11215-2806
NGC\,1386 and  NGC\,6300 the HOBL are due to a large 
contribution of intermediate age (1\,Gyr) stars. 
Intermediate age stars seem to contribute also
to the nuclear spectrum of IC\,1816, NGC\,6890 and the
radio galaxy 3\,C33.


The incidence of recent star formation seems to be related to the Hubble
type in our sample of 20  Seyfert\,2's: signatures from
young components with ages $<$500\,Myr
have been found in 4 of the 9 late-type galaxies,
but only in 2 of the 11 early-type galaxies. This tendency seems
to be present also in the sample of Gonz\'azlez Delgado et al. (1999).

For the remaining 5 late-type galaxies, two appear to present a PL
continuum (alternatively due to stellar populations younger than
10\,Myr) and three present an intermediate age (1\,Gyr) component.

Out of the nine early-type galaxies without recent central star-formation, 
the stellar population is well reproduced by the bulge template
in two of them, by the bulge plus 10\% contribution in mass of
an intermediate population in other three, while in the remaining
four, a power-law seems to be necessary. The nature of the latter component
requires further investigation, due to its degeneracy
with very young ($\le$10\,Myr) stellar population spectra
in the wavelength range investigated here.
If such component were due to very young stars in all cases,
then the number of  Seyfert\,2 with significant contribution from
young stars to the nuclear spectra would increase to
12 (60\%), in our sample of 20  Seyfert\,2's.

It is essential to continue the present investigation
along two lines: (1) assess the statistical significance of our findings
observing a larger and well defined sample of  Seyfert\,2's, together
with a comparison sample of normal galaxies of the same Hubble types;  
(2) investigate the nature of the power-law continuum.
The degeneracy of this continuum with those from very young stellar populations
could be raised, in principle, in the UV spectral region.

\acknowledgments

We thank the referee, Rosa Gonz\'alez Delgado, for many suggestions
which improved the paper. This work has benefited
also from discussions with Roberto Cid
Fernandes, Henrique Schmitt and Tim Heckman. 
We thank the support from the brazilian institutions CNPq, CAPES and
FAPERGS.

%

%
%

\begin{figure}
\plotone{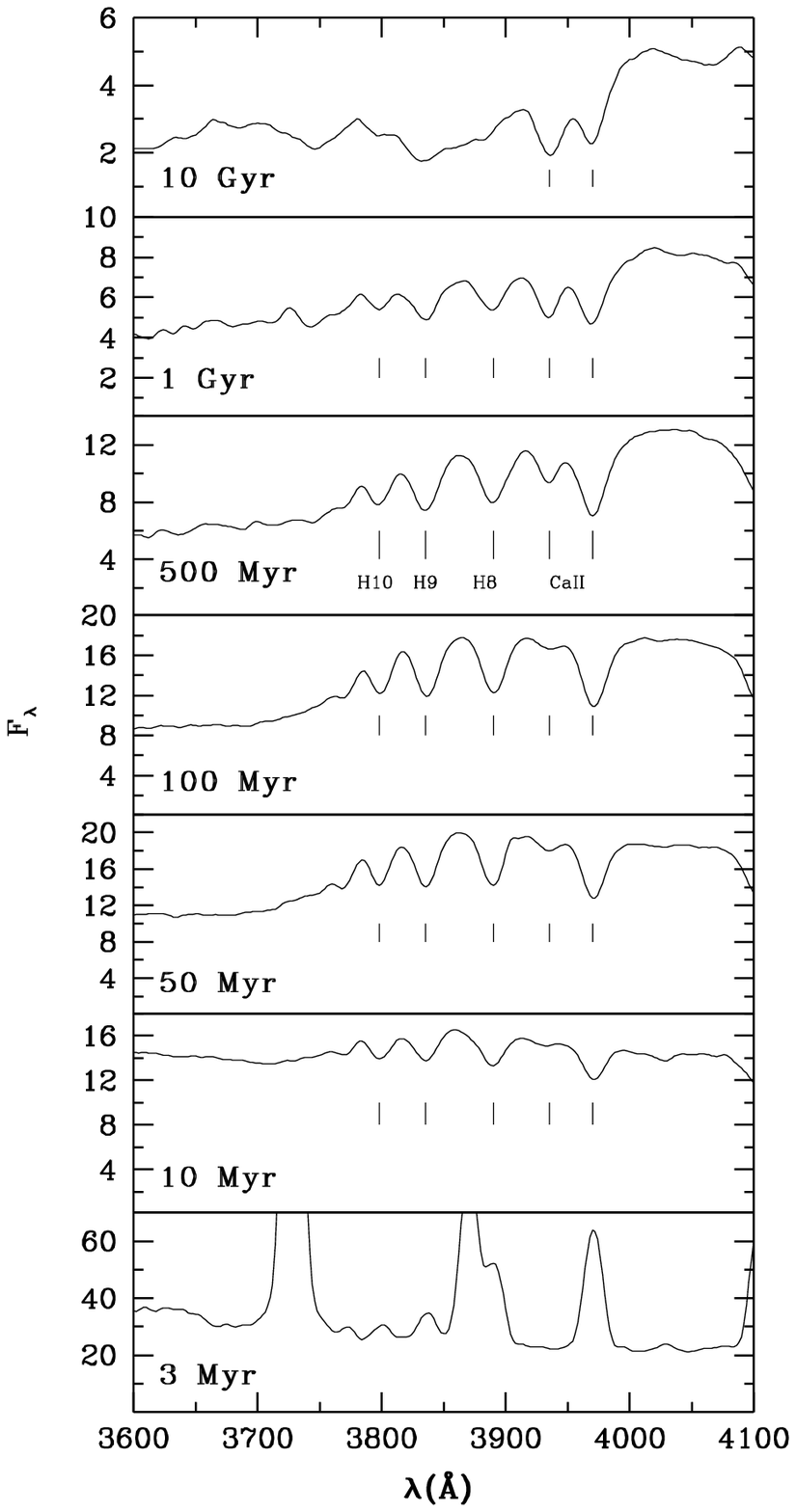}
\caption{Stellar population templates, corresponding to ages
from 3\,Myr to 10\,Gyr. The main absorption features are 
identified. \label{fig1}}
\end{figure}

\begin{figure}
\plotone{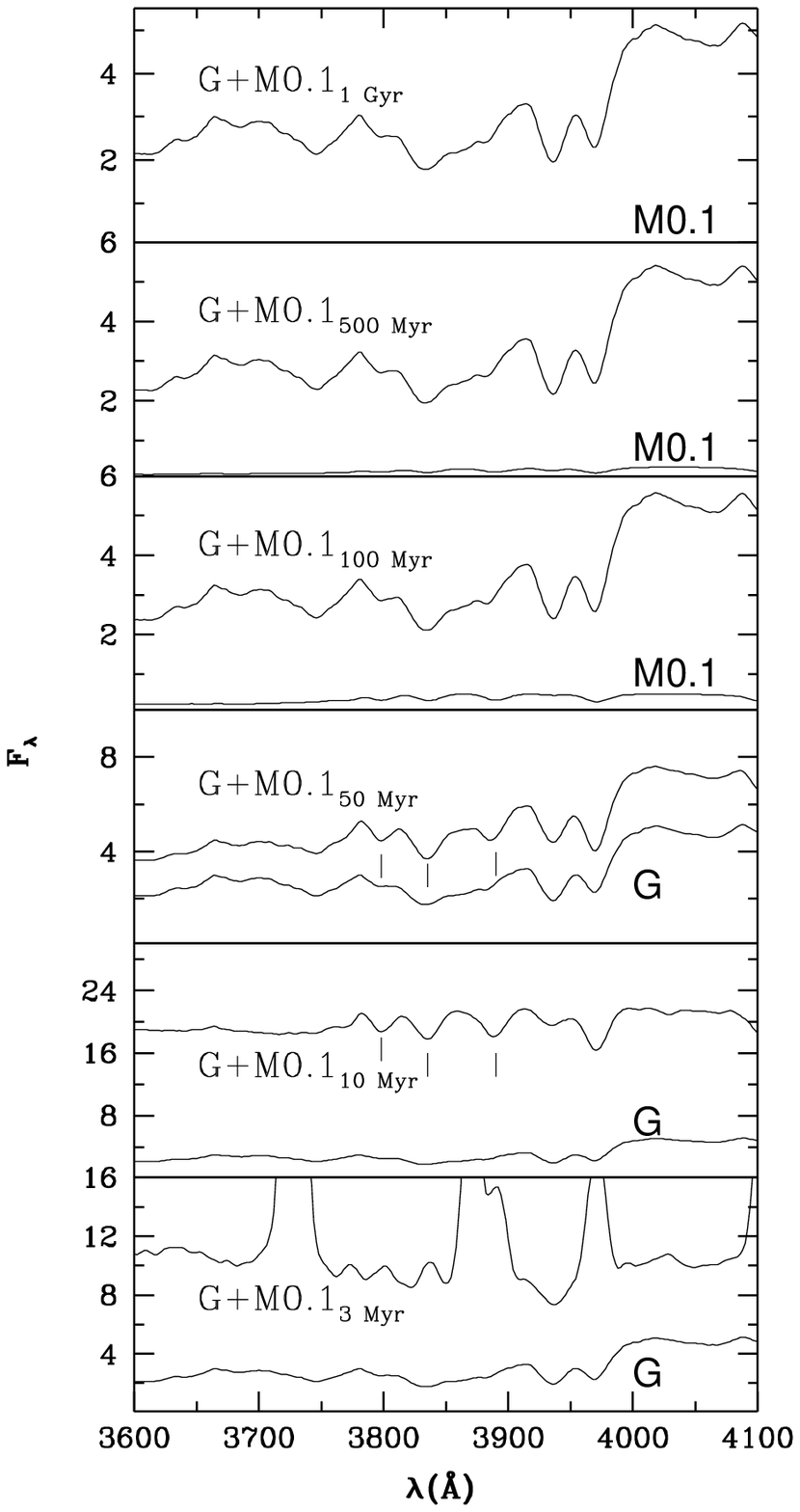}
\caption{Top of each panel: Composite spectra constructed by combining a bulge template (G) plus 0.1\% in mass (M) of bursts with ages from 3\,Myr to 1\,Gyr.
Bottom of each panel: contribution of the component with less flux to the combined template. \label{fig2}  }
\end{figure}

\begin{figure}
\plotone{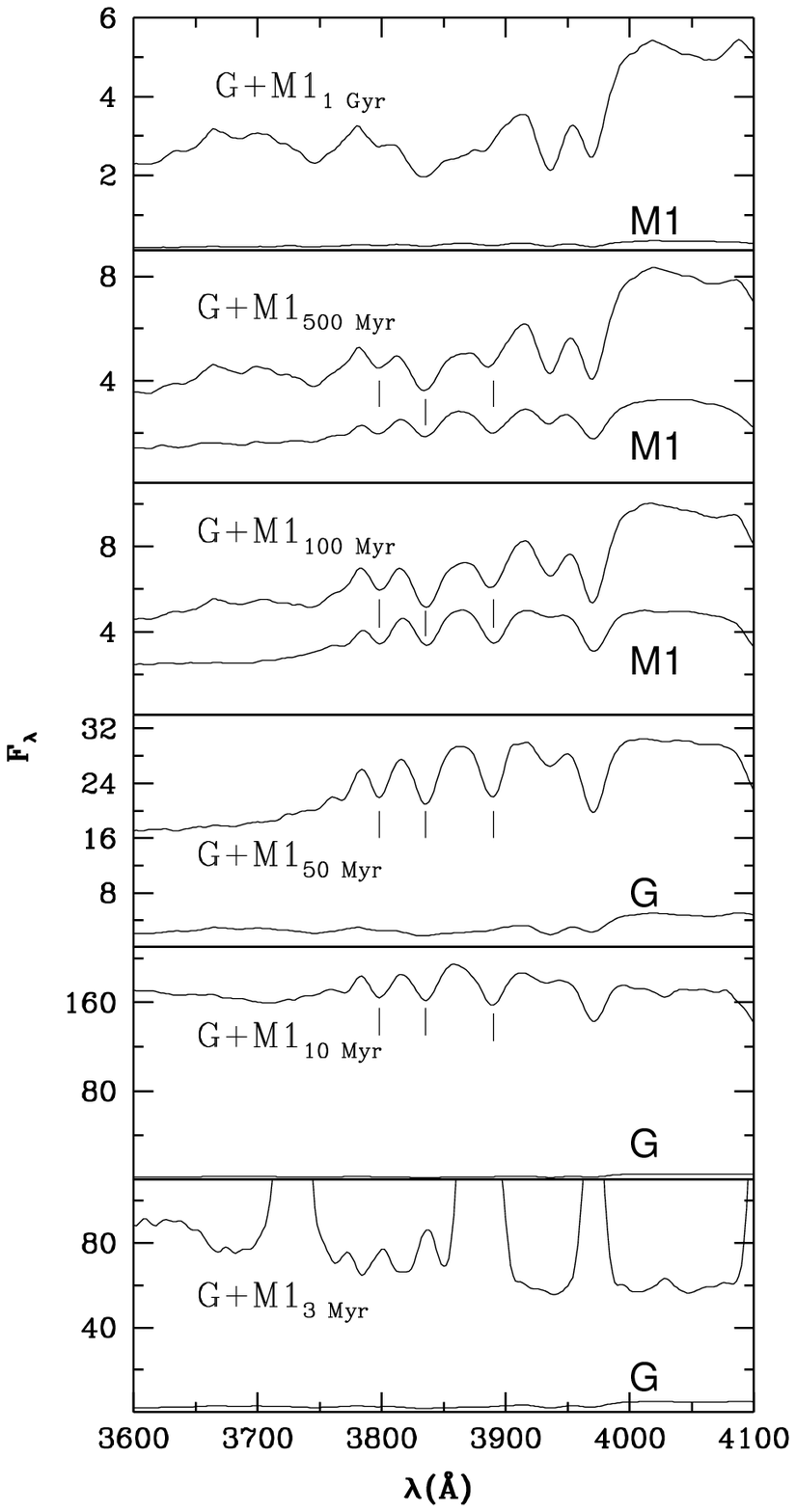}
\caption{Top of each panel: Composite spectrum constructed by combining a bulge template (G) plus 1\% in mass (M) of bursts with ages from 3\,Myr to 1\,Gyr. Bottom of each panel: contribution of the component with less flux to the combined template. \label{fig3}}
\end{figure}

\begin{figure}
\plotone{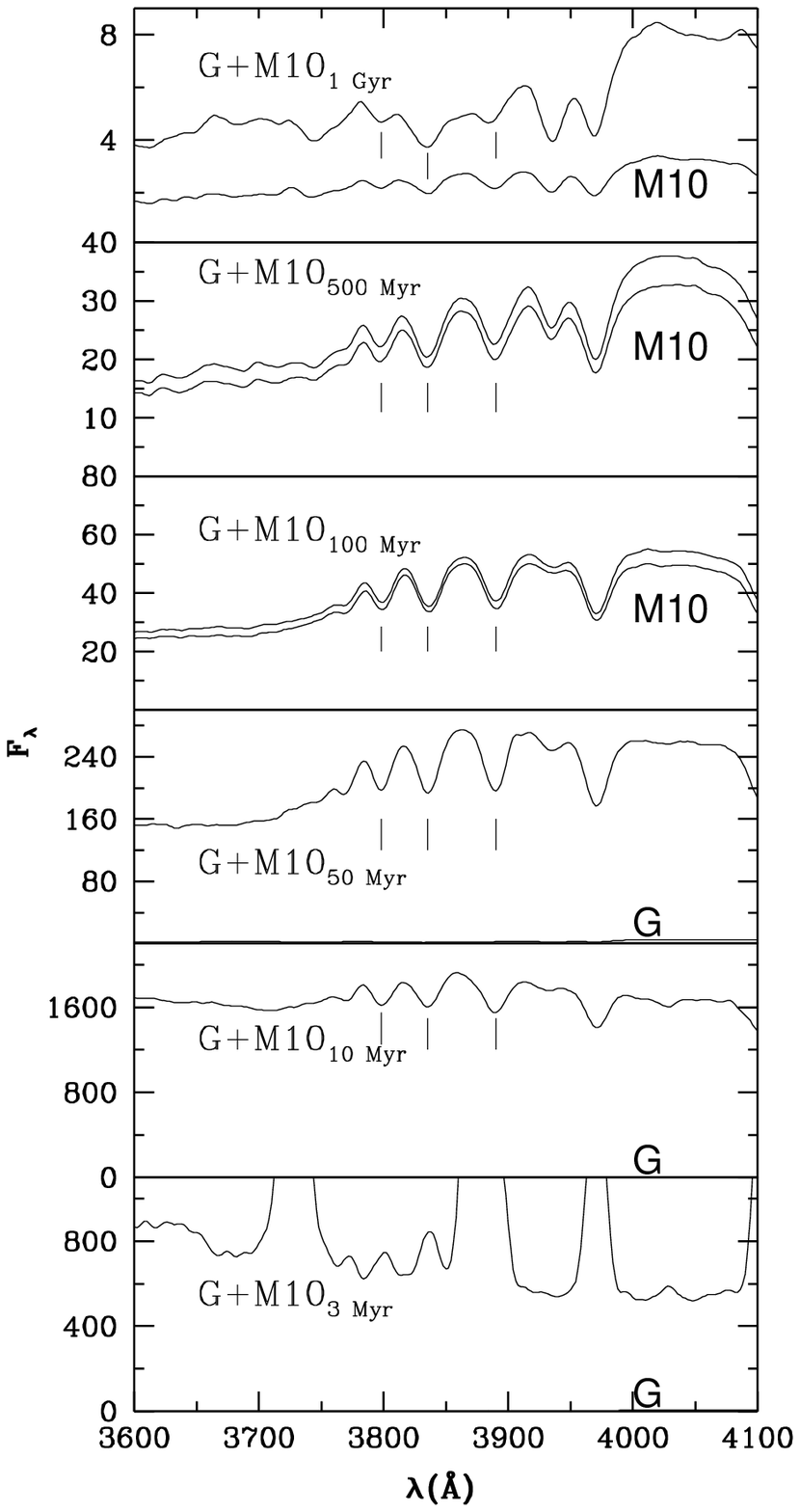}
\caption{Top of each panel: Composite spectra constructed by combining a 
bulge template (G) plus 10\% in mass (M) of bursts with ages from 
3\,Myr to 1\,Gyr. Bottom of each panel: contribution of the component with less flux to the combined template. \label{fig4}}
\end{figure}

\begin{figure}
\plotone{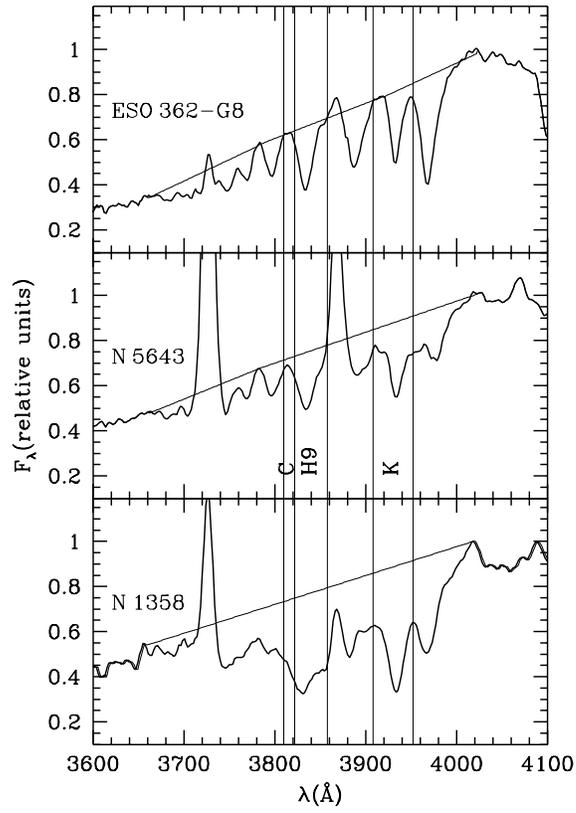}
\caption{Illustration of the continuum and windows (vertical lines) 
used in the measurements of the equivalent widths W$_C$, 
W$_{H9}$ and W$_{Ca K}$ for three  Seyfert\,2's of the sample. 
\label{fig5}}
\end{figure}

\begin{figure}
\plotone{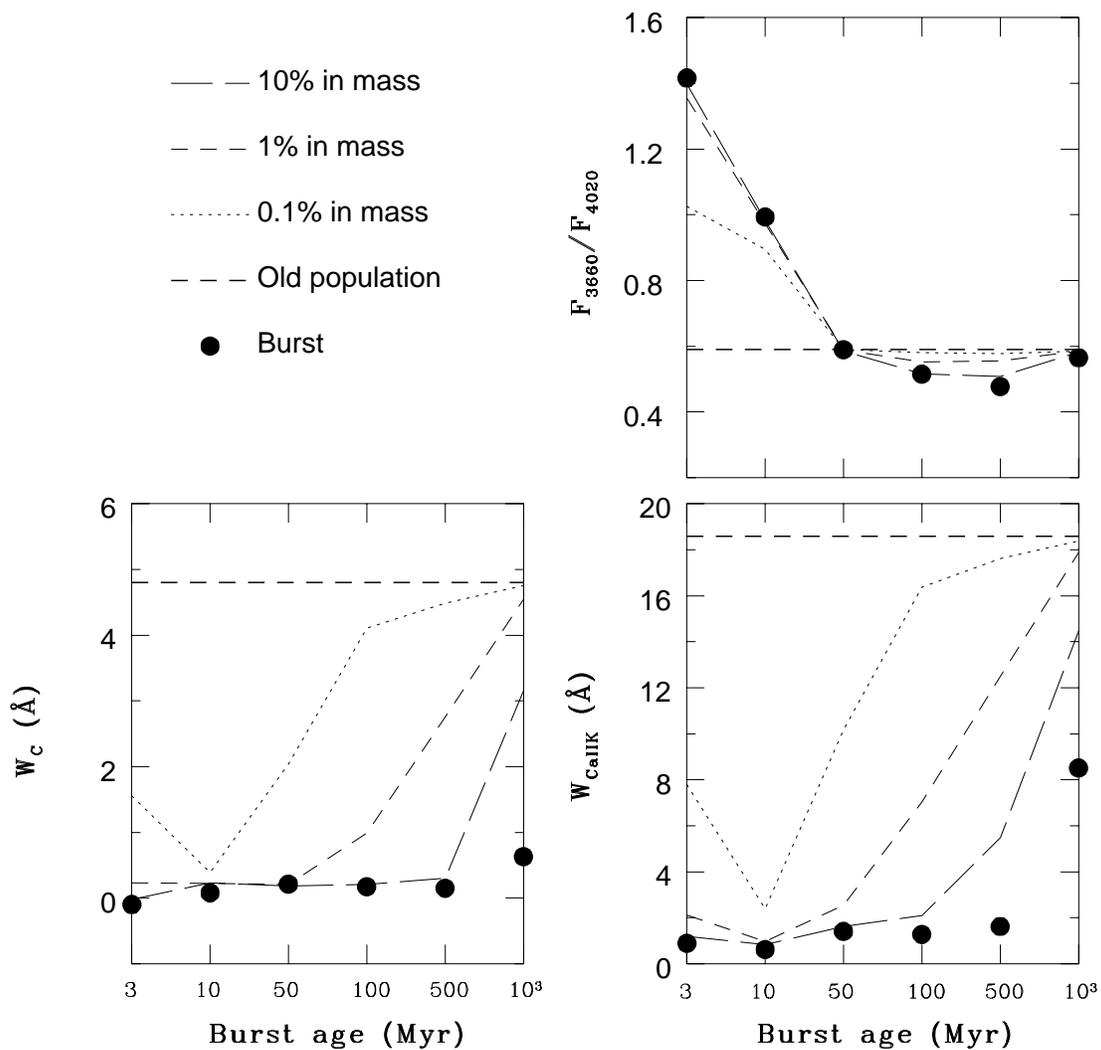}
\caption{The continuum ratio CR=F$_{3660}/F_{4020}$ and 
equivalent widths  W$_C$ and W$_{CaII\,K}$ for
synthetic spectra constructed combining a bulge template and bursts
of varying ages contributing with zero (heavy dashed line), 0.1\%
(dotted line), 1\%(dashed line), 10\% (long-dashed) and 100\%
(filled circles) in mass.\label{fig6}}
\end{figure}

\begin{figure}
\plotone{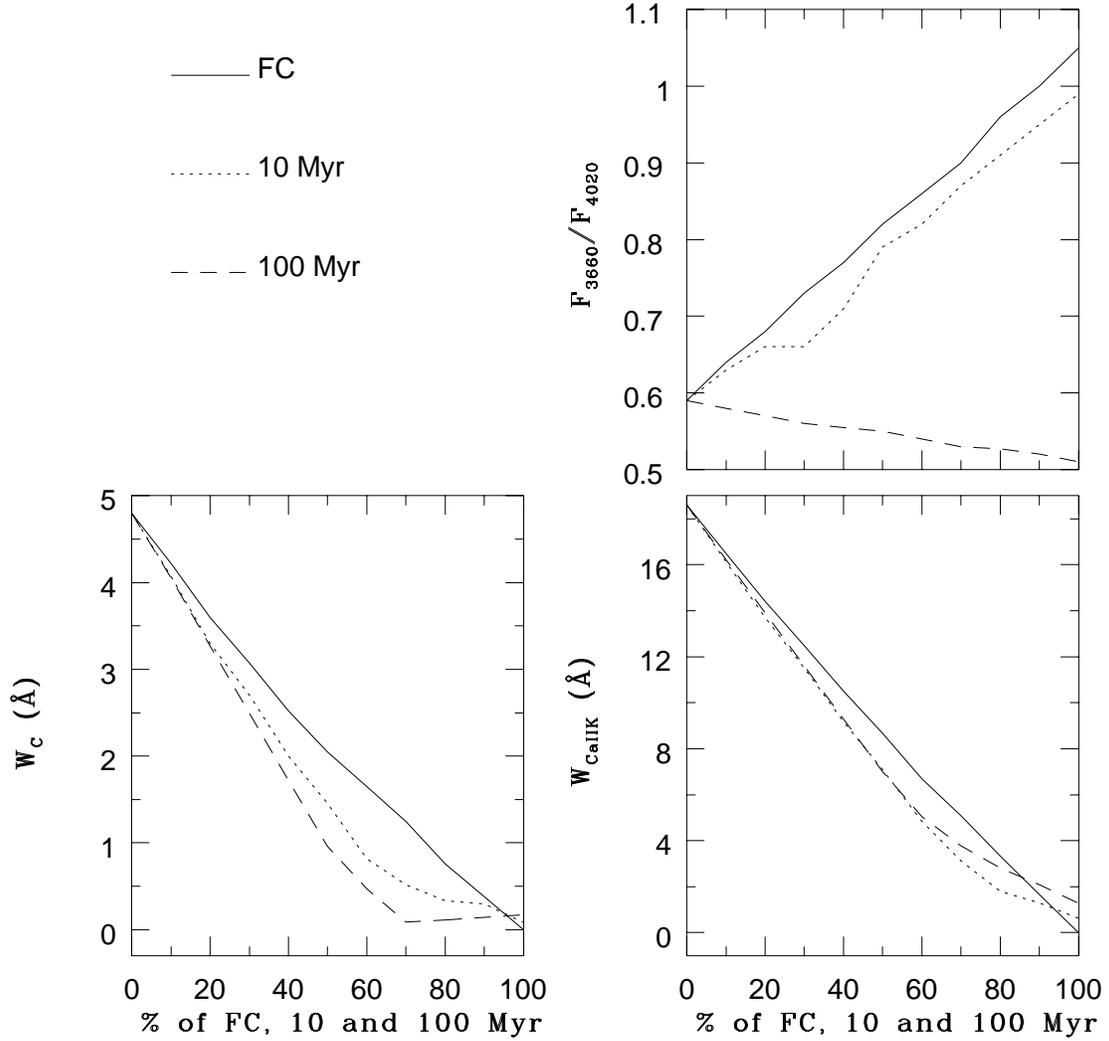}
\caption{The continuum ratio CR=F$_{3660}/F_{4020}$ and 
equivalent widths  W$_C$ and W$_{CaII\,K}$ for
synthetic spectra constructed combining a bulge template and
increasing fractions (in flux at $\lambda4020$\AA) of a PL (continuous
line), a 10\,Myr (dotted line) and a 100\,Myr (dashed line) burst
template.\label{fig7}}
\end{figure}

\begin{figure}
\plotone{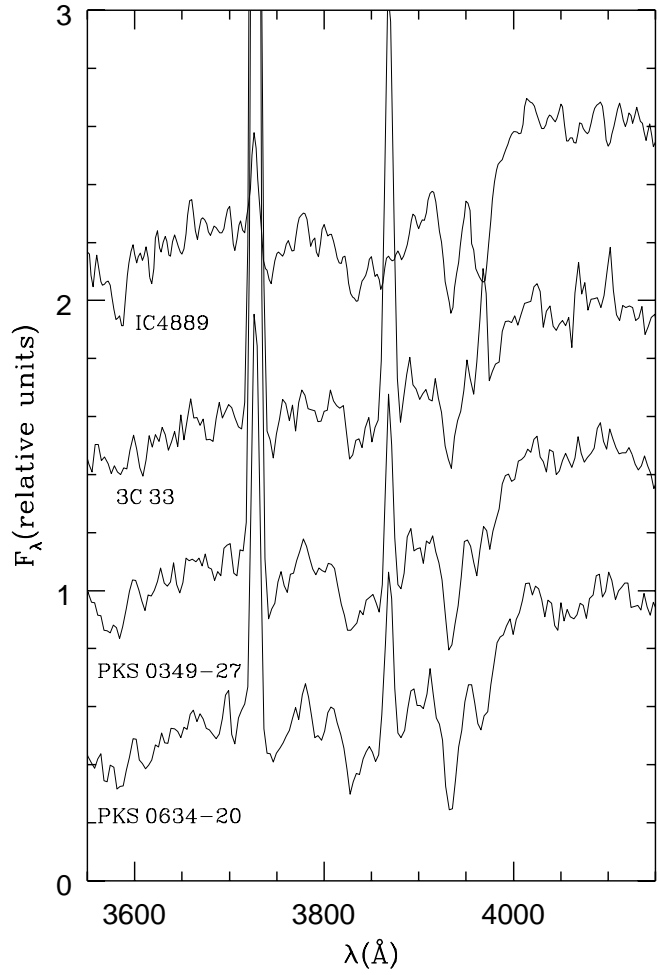}
\caption{Spectra of the elliptical and three radio galaxies
of the sample normalized at $\lambda$4020 and shifted for clarity.
\label{fig8}}
\end{figure}

\begin{figure}
\plotone{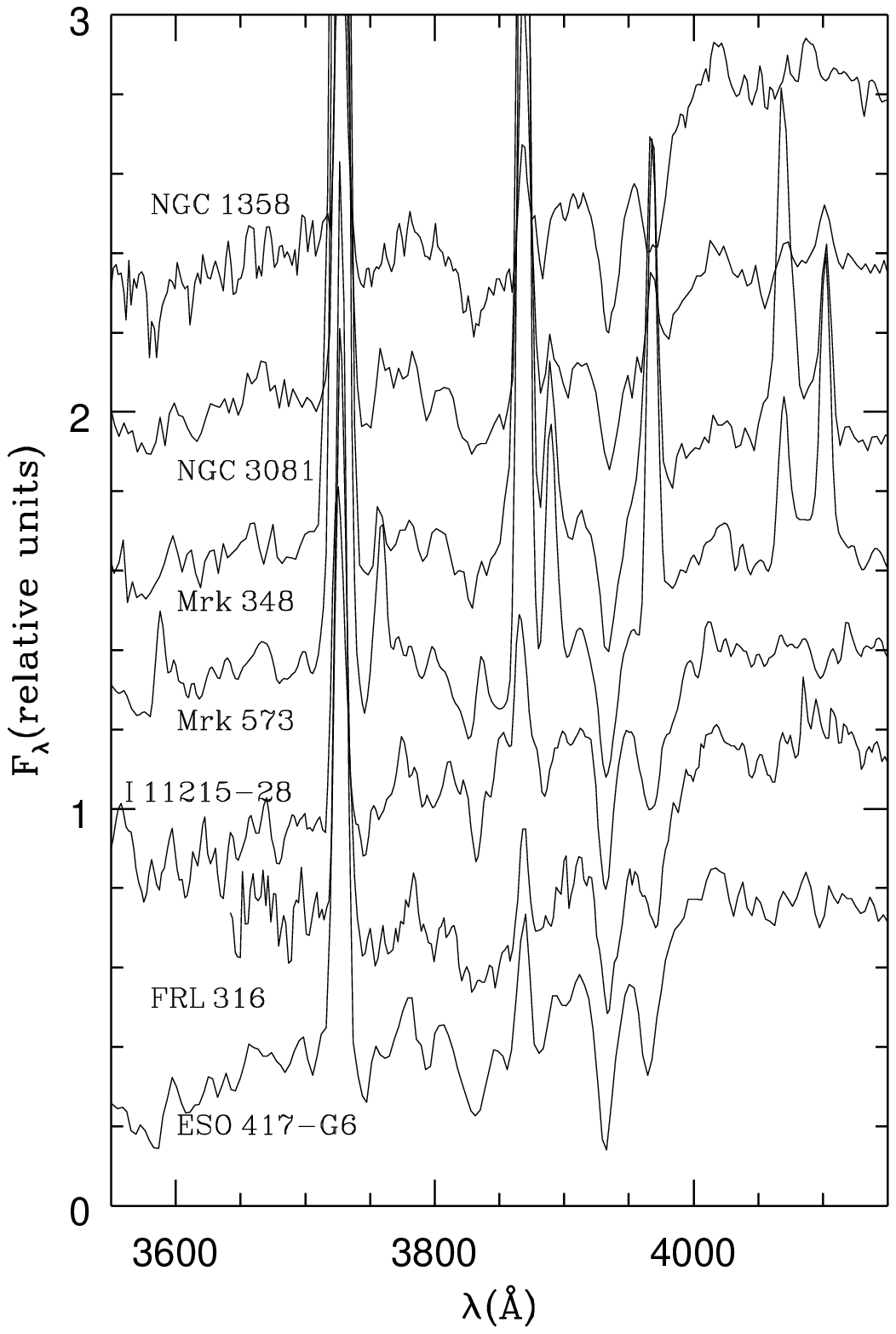}
\caption{Spectra of the 7 S0 Seyfert\,2 galaxies
of the sample normalized at $\lambda$4020 and shifted for clarity.
\label{fig9}}
\end{figure}

\begin{figure}
\plotone{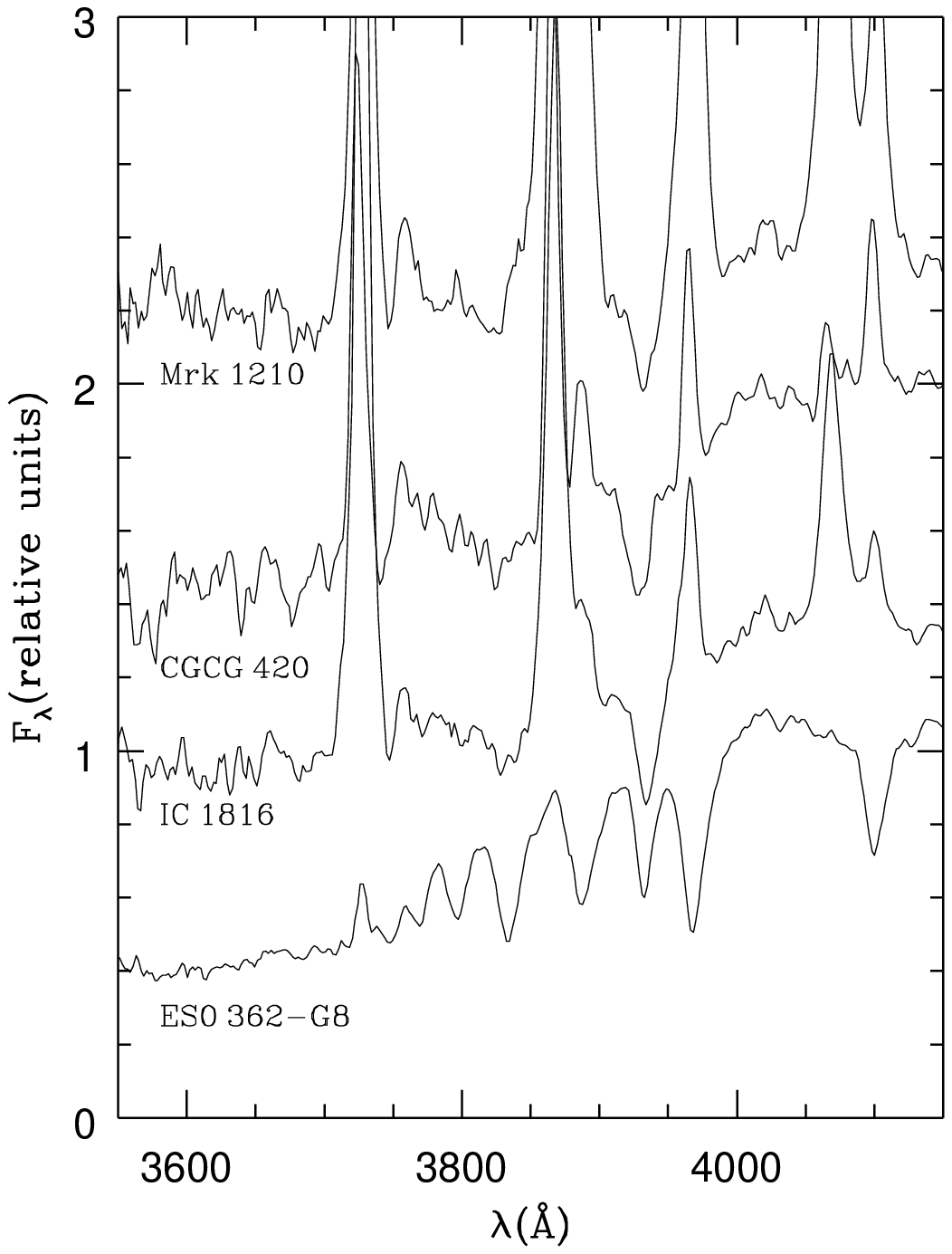}
\caption{Spectra of the 4 Sa Seyfert\,2 galaxies
of the sample normalized at $\lambda$4020 and shifted for clarity.
\label{fig10}}
\end{figure}

\begin{figure}
\plotone{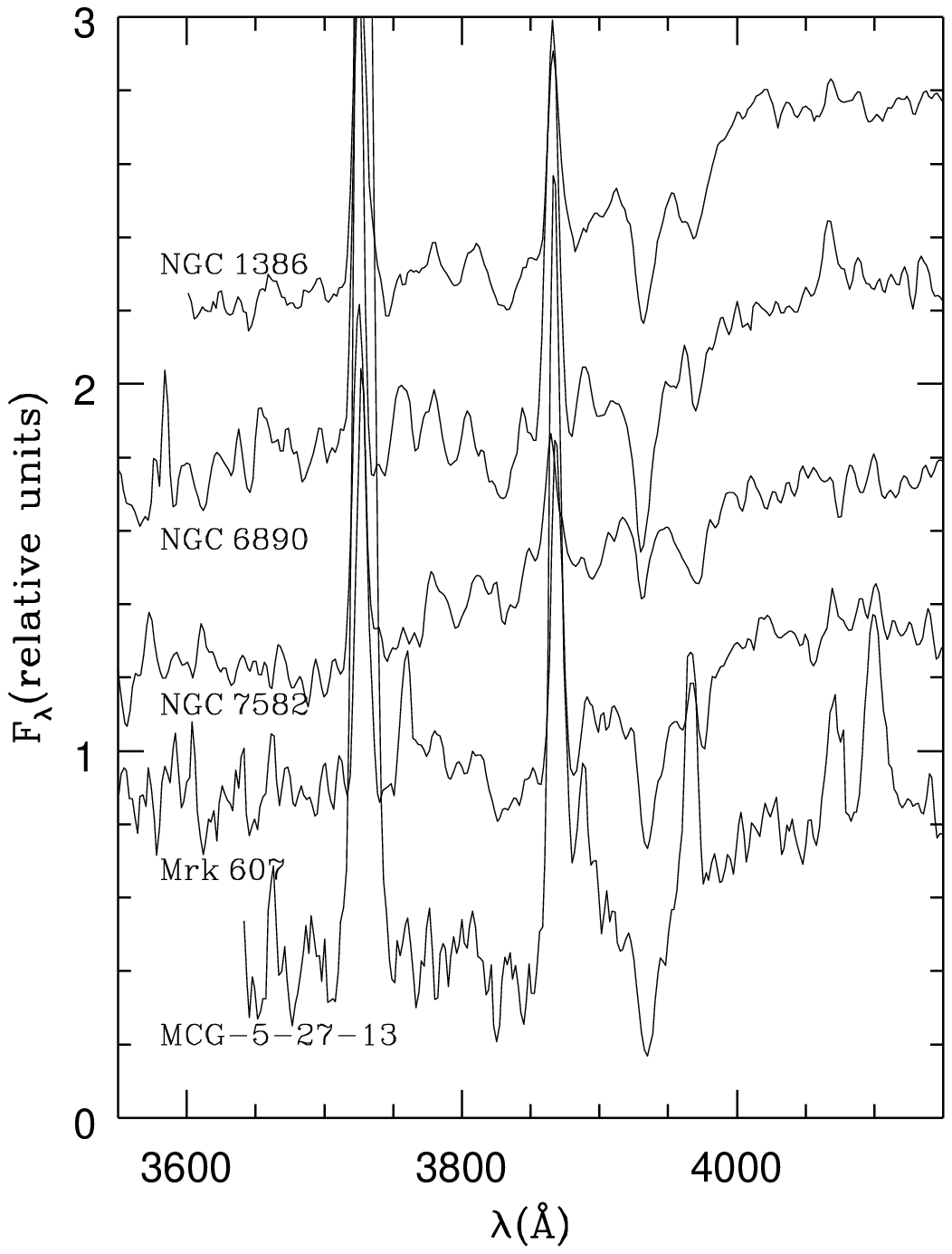}
\caption{Spectra of the 5 Sb Seyfert\,2 galaxies
of the sample normalized at $\lambda$4020 and shifted for clarity.
\label{fig11}}
\end{figure}

\begin{figure}
\plotone{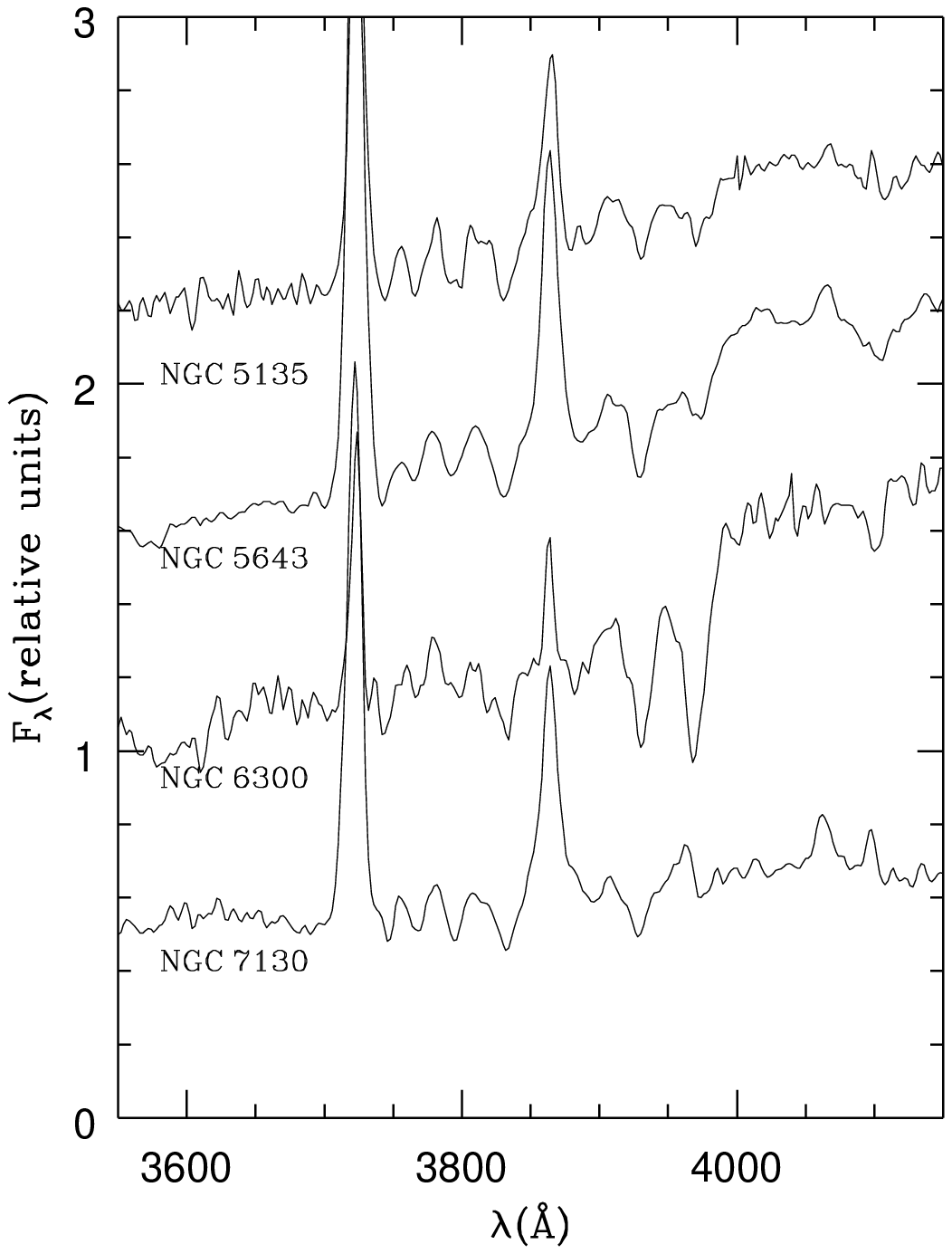}
\caption{Spectra of the 4 Sc Seyfert\,2 galaxies
of the sample normalized at $\lambda$4020 and shifted for clarity.
\label{fig12}}
\end{figure}

\begin{figure}
\plotone{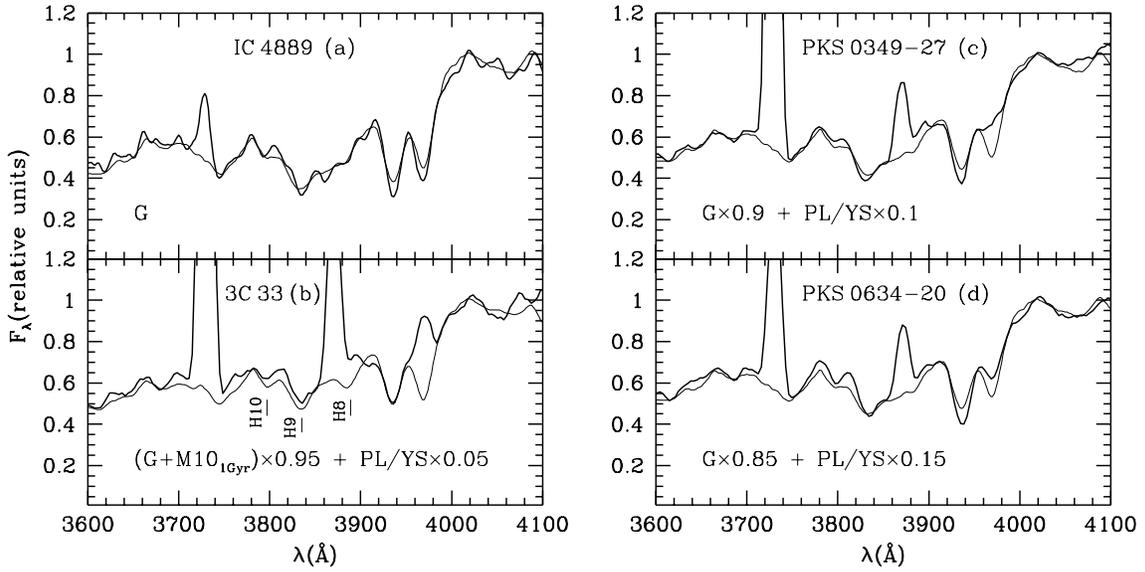}
\caption{Comparison of the nuclear spectrum of the normal elliptical 
galaxy IC\,4889 and the three radio ellipticals (heavy lines) 
with the best models (thin lines). Labels: G represents the bulge template,
M$x_y$ represents a burst of age $y$ contributing with $x$\% (in mass)
of the bulge mass and PL/YS is the power-law/young stars component. 
The high order Balmer lines H10, H9 and
H8 (HOBL) are identified by vertical lines. \label{fig13}}
\end{figure}

\begin{figure}
\plotone{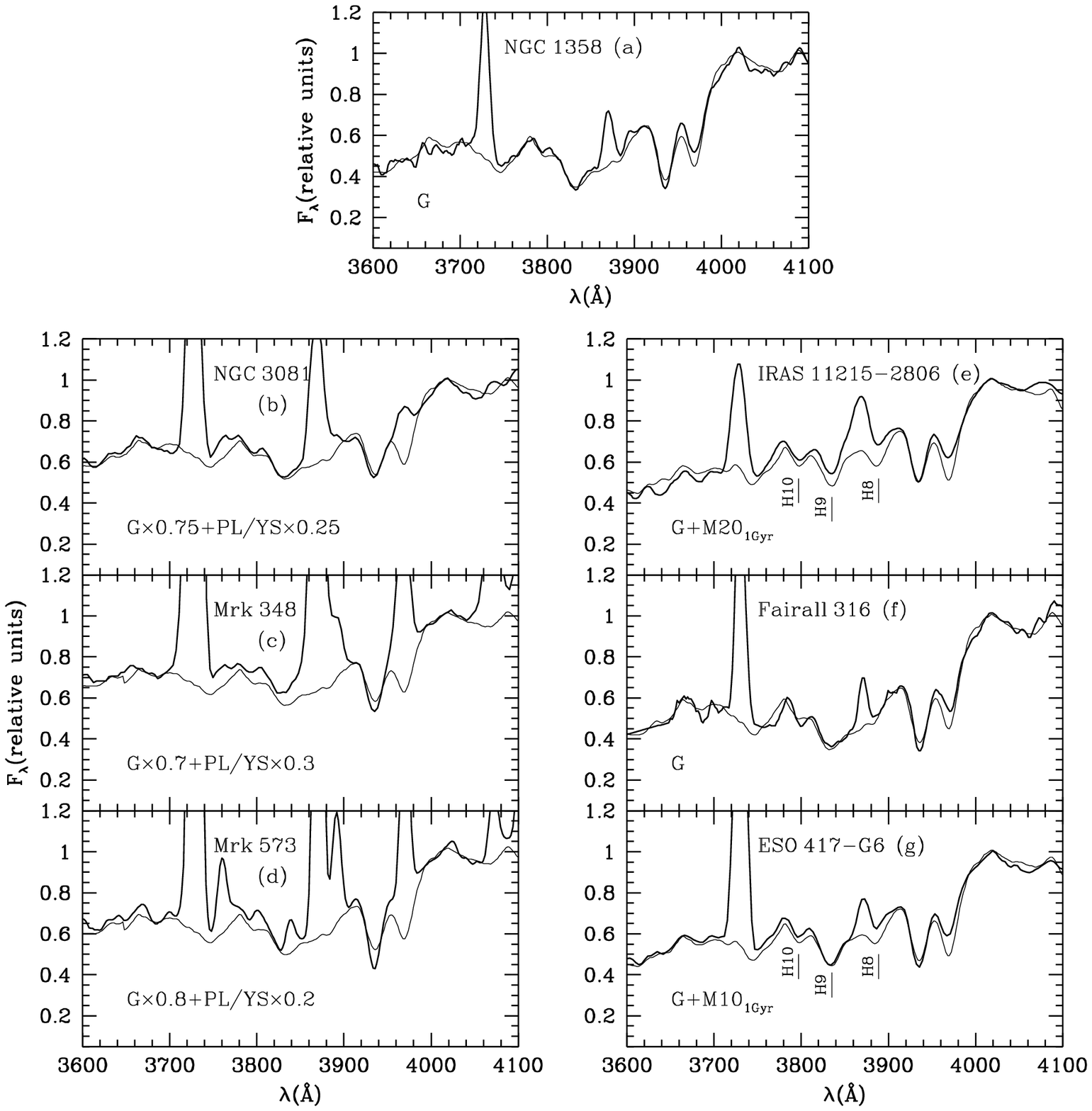}
\caption{Comparison between the nuclear spectrum of the S0 Seyfert's
(heavy lines) and the best models (thin lines). Labels as in
Fig.\ref{fig13}. Models for Mrk\,348 and Mrk\,573 include the 
Balmer continuum from the emitting gas.\label{fig14}}
\end{figure}

\begin{figure}
\plotone{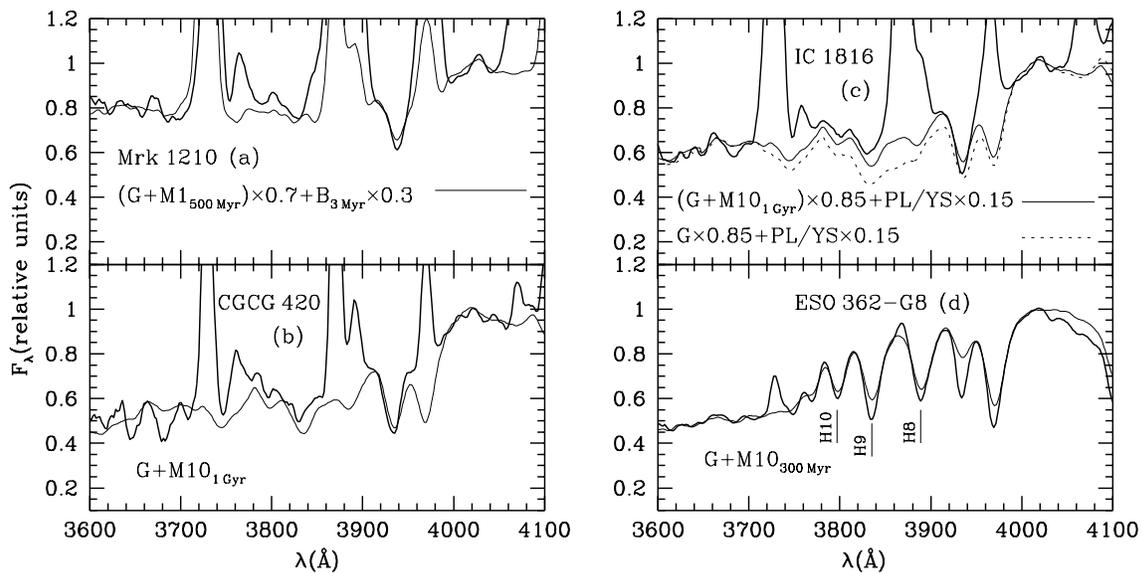}
\caption{Comparison between the nuclear spectrum of the Sa Seyfert's 
(heavy lines) and the best models (thin lines). Most labels as in
Fig.\ref{fig13} with B$_{3\,Myr}$ representing the 3\,Myr burst.
Models for Mrk\,1210 and IC\,1816 include the Balmer continuum from the emitting gas.\label{fig15}}
\end{figure}

\begin{figure}
\plotone{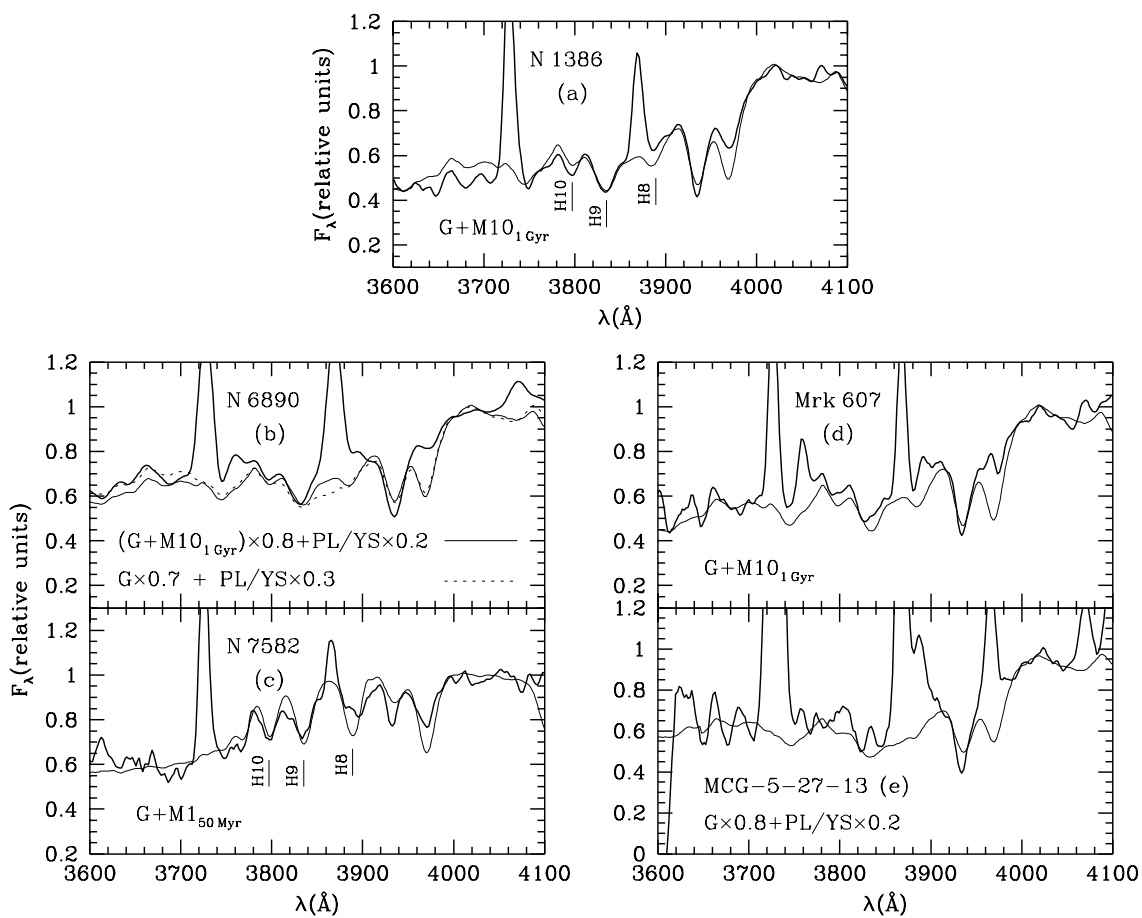}
\caption{Comparison between the nuclear spectrum of the Sb Seyfert's
(heavy lines) and the best models (thin lines). Labels as in
Fig.\ref{fig13}. Model for MCG-05-27-13 includes the 
Balmer continuum from the emitting gas.\label{fig11}}
\end{figure}

\begin{figure}
\plotone{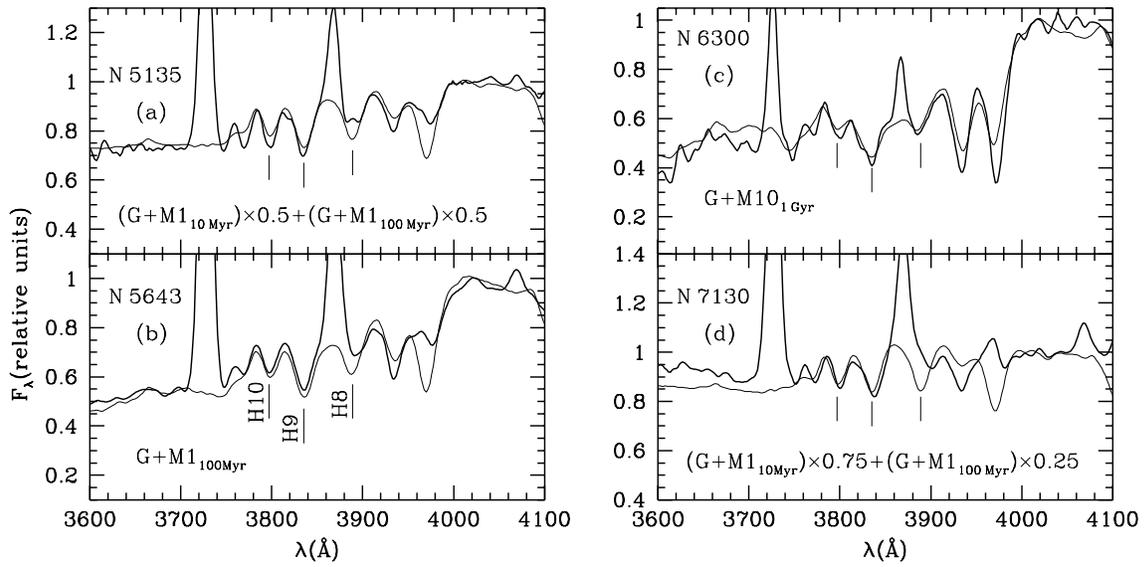}
\caption{Comparison between the nuclear spectrum of the Sc Seyfert's
(heavy lines) and the best models (thin lines). Labels as in
Fig.\ref{fig13}.\label{fig17}}
\end{figure}

\begin{figure}
\plotone{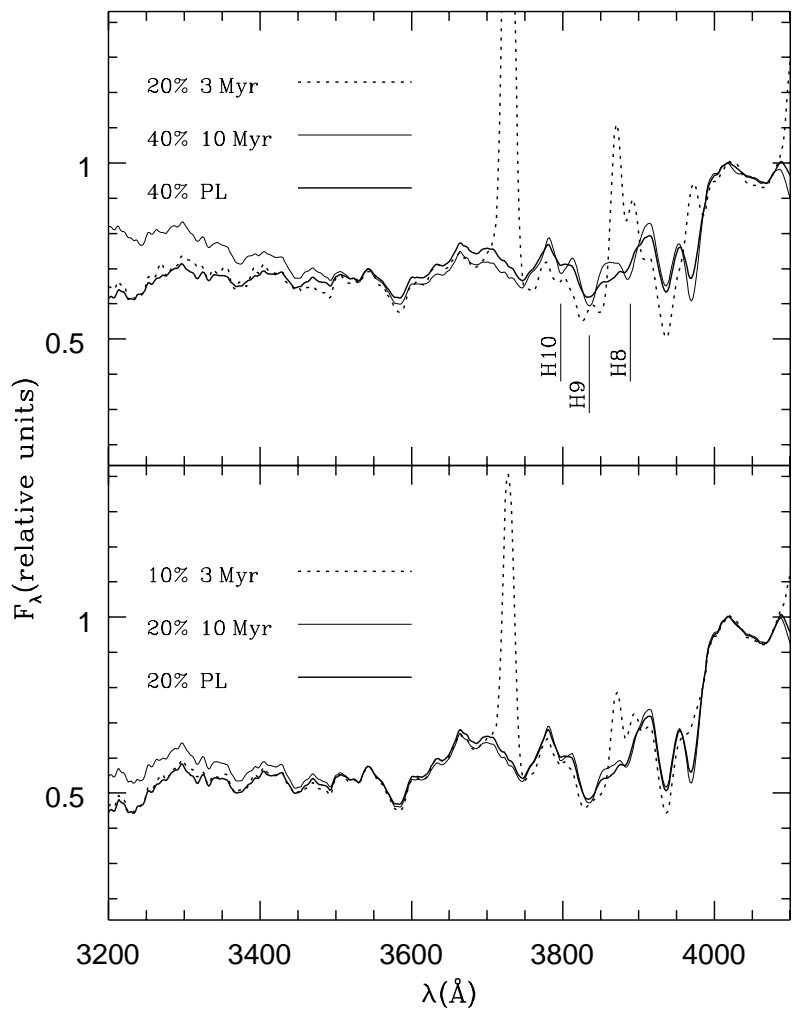}
\caption{Comparison between model spectra constructed combining
the bulge template with: (i) a PL (heavy line) contributing with
20\% (bottom panel) and 40\% (upper panel) of the flux at $\lambda$4020\AA;
(ii) a 10\,Myr population template (thin line) with the same contribution as
the PL; and (iii) a 3\,Myr population template (dotted line) contributing with
10\% (bottom panel) and 20\% (upper panel) of the flux at $\lambda$4020\AA.
\label{fig18}}
\end{figure}

%
%
\clearpage

\begin{table*}
\begin{center}
\begin{tabular}{lrr}
\tableline
Age & M/L$_V$ & L$_{\lambda 4020}$ \\
\tableline
3\,Myr &0.030 &1134.0\\
10\,Myr &0.007 &3354.1\\
50\,Myr &0.054 &504.4\\
100\,Myr &0.260 &97.1\\
500\,Myr &0.310 &63.9\\
1\,Gyr &2.19 &6.8 \\
10\,Gyr &8.6 &1.0\\
\tableline
\end{tabular}
\end{center}
\caption{Template ages, mass-to-light ratios in V and relative luminosity
at $\lambda$4020\AA}
\label{tbl-1}
\end{table*}

\clearpage

\begin{table*}
\begin{center}
\begin{tabular}{lrrrrr}
\tableline
Name& $\lambda$3660/4020& W$_C$& W$_{H9}$& W$_{CaII\,K}$& Scale \\
&&(\AA)&(\AA)&(\AA)&(pc/arcsec)\\
\tableline
NGC1358    &0.51   & 4.5 &18.5 &18.6 &257 \\
MGC1386    &0.52   & 2.5 &12.9 &14.0 & 48 \\
NGC3081    &0.69   & 3.5 &12.8 &14.0 &140 \\
NGC5135    &0.74   & 0.4 & 4.0 & 2.7 &256 \\
NGC5643    &0.55   & 0.8 & 7.4 & 9.5 & 69 \\
NGC6300    &0.54   & 3.0 &12.9 &15.8 & 63 \\
NGC6890    &0.72   & 3.0 &10.1 &13.7 &159 \\
NGC7130    &0.93   & 0.5 & 3.1 & 3.3 &314 \\
NGC7582    &0.62   & 0.7 & 4.1 & 3.7 &100 \\
Mrk348     &0.72   & 2.4 & 8.6 &13.0 &302 \\
Mrk573     &0.70   & 3.7 &11.8 &15.1 &334 \\
Mrk607     &0.63   & 2.9 &12.8 &14.3 &176 \\
Mrk1210    &0.83   & 1.6 &em   & 9.5 &253 \\
CGCG420-015 &0.57  & 2.6 & 9.8 &13.8 &570 \\
IC1816      &0.66  & 2.2 & 6.9 &12.8 &328 \\
IRAS11215-2806 &0.55 & 1.4 & 7.4 &11.8 &262 \\
MCG-05-27-013 &0.67 & 2.9 &11.6 &15.1 &470 \\
Fairall316  &0.57  & 4.6 &17.9 &18.2 &308 \\
ESO417-G6   &0.59  & 3.5 &13.3 &15.2 &310 \\
ESO362-G8   &0.52  & 0.1 & 7.4 & 5.8 &298 \\
3C33        &0.60  & 2.4 &12.9 &15.2 &1114 \\
P0349-27    &0.60  & 4.4 &16.1 &18.5 &1240 \\
P0634-20    &0.66  & 3.4 &16.2 &17.9 &1055 \\
\tableline
\end{tabular}
\end{center}
\caption{Near-UV measurements and scale of the Sample}
\label{tbl-2}
\end{table*}

\clearpage


\begin{table*}
\begin{center}
\begin{tabular}{lrr}
\tableline
Galaxy&RC3 &Malkan et al.\\
\tableline
Mrk\,607       &Sa          &Sb\\
NGC\,1386      &SB0         &Sbc\\
CGCG\,420-015  &?           &Sa\\
ESO\,362-G8     &S0         &Sa\\
Mrk\,1210       &S?         &Sa\\
IRAS\,11215-2806  &?        &S0\\
MCG-05-27-013   &SBa       &Sb\\
NGC\,5135         &SBab      &Sc\\
NGC\,6300         &SBb       &Sd\\
NGC\,7130         &Sa        &Sd\\
\tableline
\end{tabular}
\end{center}  
\caption{Previous and new classifications\label{tbl-3}}
\end{table*}


\begin{references}

\reference {} Bica, E. \& Alloin, D. 1986, A\&A, 162, 21

\reference {} Bica, E. \&  Alloin, D. 1987, A\&AS, 70, 281


\reference {} Bica, E., Alloin, D. \& Schmidt, A. 1990, \mnras, 242, 241

\reference {} Bica, E., Alloin, D. \& Schmitt, H. R. 1994, A\&A, 283, 805

\reference {} Cid Fernandes, R. \& Terlevich, R. 1992, in {\it Relationships
between Active Galactic Nuclei and Starbursts Galaxies}, ASP Conf. Ser., 31, Astron. Soc. Pac. (San Francisco)

\reference {} Cid Fernandes, R. \& Terlevich, R. 1995, \mnras, 272, 423

\reference {} Cid Fernandes, R., Schmitt, H. R. \& Storchi-Bergmann, T. 1998,
\mnras, 297, 579

\reference {} Cid Fernandes, R., Lacerda, R. R., Schmitt, H. R. \& Storchi-Bergmann, T. 1999, IAU Symp. 193, eds.  K. A. van der Hucht, G. Koenigsberger \& P. R. J. Enens, ASP, p.590

\reference {} de Vaucouleurs, G. et al. 1991, Third Reference Catalog of Bright
Galaxies (RC3), New York: Springer

\reference {} Fraquelli, H. A., Storchi-Bergmann, T. \& Binette, L. 2000,
\apj, 532, 867

\reference {} Gonz\'alez Delgado, R. M., Heckman, T. M. \& Leitherer, C. 2000,
in press

\reference {} Gonz\'alez Delgado, R. M., Leitherer, C. \& Heckman, T. M. 1999,
\apjs, 125, 489 

\reference {} Gonz\'alez Delgado, R. M., Heckman, T., Leitherer, C.,
Meurer, G., Krolik, J., Wilson, A. S., Kinney, A. \& Koratkar, A.
1998, \apj, 505, 174

\reference {} Heckman, T., Krolik, J., Meurer, G., Calzetti, D., Kinney, A,
Koratkar, A., Leitherer, C., Robert, C. \& Wilson, A. S. 1995, \apj, 452, 549

\reference {} Heckman, T. M., Gonz\'alez Delgado, R., Leitherer, C., Meurer, G. R., Krolik, J.,
Wilson, A. S., Koratkar, A. \& Kinney, A.  1997, \apj, 482, 114 (H97)

\reference {} Leitherer, C. et al. 1996, PASP, 108, 996

\reference {} Malkan, M. A., Gorjian, V. \& Tam, R. 1998, \apjs, 117, 25


\reference {} Norman, C. \& Scoville, N. 1988, \apj, 332, 124

\reference {} Perry, J. J, \& Dyson, J. E. 1985, \mnras, 213, 665

\reference {} Pogge, R. W. \& De Robertis, M. M. 1993, \apj, 404, 563

\reference {} Schmidt. A. A., Alloin, D. \& Bica, E., 1995, MNRAS, 273, 945

\mnras, 278, 965


\reference {} Schmitt, H. R., Storchi-Bergmann, T. \& Cid Fernandes, R. 1999,
\mnras, 304, 35

\reference {} Seaton, M J. 1979, \mnras, 187, 73P

 
\reference {} Storchi-Bergmann, T., Bica, E., Kinney, A. L. \& Bonatto, C. 1997
 MNRAS, 290, 231

\reference {} Storchi-Bergmann, T., Cid Fernandes, R. \& Schmitt, H. R. 1998,
\apj, 501, 94

\reference {} Storchi-Bergmann, T., Kinney, A. \& Challis, P. 1995, \apjs,
98, 103

\reference {} Terlevich, E., Diaz, A. I. \& Terlevich, R. 1990, \mnras, 242, 271

\reference {} Tran, H. D. 1995a, \apj, 440, 565

\reference {} Tran, H. D. 1995b, \apj, 440, 578

\reference {} Tran, H. D. 1995c, \apj, 440, 597



\end{references}
\end{document}